%% file: 1loop.tex
\documentclass[11pt]{article}
\usepackage[left=1in,right=1in]{geometry}
\usepackage{hepth}
\usepackage{empheq}
\usepackage[pdfstartview=FitH,pdfpagemode=None]{hyperref}
\input{Definitions}

\begin{document}
\input{Cover}
\tableofcontents
\newpage
\input{Introduction}

\input{Review_Bulk}
\input{Review_D5}
\input{Fluctuations_Bosons}
\input{Fluctuations_Fermions}

\input{Fluctuations_EOMbosons}
\input{Fluctuations_EOMfermions}
\input{Fluctuations_WL}

\input{Determinant_Zeta}

\input{Determinant_Modes}
\input{Determinant_Kernelbosons_Triplet}
\input{Determinant_Kernelbosons_Transverse}
\input{Determinant_Kernelbosons_Mixed}
\input{Determinant_Kernelbosons_Special}
\input{Determinant_Kernelbosons_All}
\input{Determinant_Kernelfermions}
\input{Determinant_Kernelfinal}

\input{Conclusions}

\appendix
\input{Appendix_Conventions}

\input{Appendix_Embeddings}
\input{Appendix_AdS2}
\input{Appendix_Calcsigma}
\bibliographystyle{JHEP}
\bibliography{det_sec2}
\end{document}

%% file: Definitions.tex
\numberwithin{equation}{section}

\newcommand{\ie}{i.e.,\ }

\newcommand{\rmd}{\,\mathrm{d}}

\newcommand{\Tr}{\operatorname{tr}}

\newcommand{\idmat}{\mathbb{I}}
\newcommand{\Itwo}{\idmat_2}
\newcommand{\Ifour}{\idmat_4}

\newcommand{\flatind}[1]{{\underline{#1}}}

\newcommand{\im}{\operatorname{Im}}

\newcommand{\e}[1]{\operatorname{e}^{#1}}

\newcommand{\rh}{r_+}

\newcommand{\Sb}{I_{\text{bound}}}

\newcommand{\prefac}{\frac{N}{3\pi^2 \alpha'}}

\newcommand{\Lag}{\mathcal{L}}

\newcommand{\F}{\mathcal{F}}
\newcommand{\A}{\mathcal{A}}

\newcommand{\lad}{L}

\newcommand{\Order}{\mathcal{O}}

\newcommand{\res}{\operatorname{Res}}

\newcommand{\vt}{\vartheta}

\newcommand{\bt}{{\bar{t}}}

\newcommand{\unitAds}{{\widehat{AdS}_2}}
\newcommand{\unitS}{{\widehat{S}^4}}

%% file: Cover.tex
\preprint{MCTP-11-43}
\title{{\bf One-loop Effective Action of the Holographic Antisymmetric Wilson Loop}}

\authors{%
Alberto Faraggi$^a$, Wolfgang M\"uck$^{b,c}$ and Leopoldo A. Pando Zayas$^a$\\
~\\~\\
\small {\it%
{}$^a$Michigan Center for Theoretical Physics \\
Randall Laboratory of Physics, The University of Michigan \\
Ann Arbor, MI 48109-1040, USA\\
~\\
{}$^b$Dipartimento di Scienze Fisiche, Universit\`a degli Studi di Napoli ``Federico II"\\
Via Cintia, 80126 Napoli, Italy\\
~\\
{}$^c$INFN, Sezione di Napoli, Via Cintia, 80126 Napoli, Italy\\
}~\\~\\
{\texttt{faraggi@umich.edu, wolfgang.mueck@na.infn.it, lpandoz@umich.edu}}
}

\date{}

\abstract{%
We systematically study the spectrum of excitations and the one-loop determinant of holographic Wilson loop operators in antisymmetric representations of $\mathcal{N}=4$ supersymmetric Yang-Mills theory.
Holographically, these operators are described by D5-branes carrying electric flux and wrapping an $S^4 \subset S^5$ in the $AdS_5\times S^5$ bulk background. We derive the dynamics of both bosonic and fermionic excitations for such D5-branes. A particularly important configuration in this class is the
D5-brane with $AdS_2\times S^4$ worldvolume and $k$ units of electric flux, which is dual to the circular Wilson loop in the totally antisymmetric representation of rank $k$. For this Wilson loop, we obtain the spectrum, show explicitly that it is supersymmetric and calculate the one-loop effective action using heat kernel techniques.%
}

\maketitle 

%% file: Introduction.tex
\section{Introduction}
\label{intro}

Wilson loop operators play a central role in gauge theories, both as formal variables and as important order parameters. In the context of the AdS/CFT correspondence expectation values of Wilson loops were first formulated by Maldacena \cite{Maldacena:1998im} and Rey-Yee \cite{Rey:1998ik}.

One of the most exciting developments early on was the realization that the expectation value of the BPS circular Wilson loop can be computed using a Gaussian matrix model \cite{Erickson:2000af, Drukker:2000rr}. This conjecture was later rigorously proved in \cite{Pestun2009}. In a beautiful, now classic work by Gross and Drukker, the matrix model was evaluated and its leading $N$, large 't~Hooft coupling limit was successfully compared with the string theory answer. One of the most intriguing windows opened by this problem is the question of quantum corrections it their entire variety. For example, having an exact field theory answer (Gaussian matrix model) prompted Gross and Drukker to speculate that the exact matrix model result was the key to understanding higher genera on the string theory side. The quantum corrections on the string theory side have been the subject of much investigation starting with earlier efforts in \cite{Forste:1999qn, Drukker:2000ep}  and continuing in more recent works such as \cite{Sakaguchi:2007ea, Kruczenski:2008zk}. Despite these concerted efforts, it is fair to say that a crisp picture of matching the BPS Wilson loop at the quantum level on both sides of the correspondence has not yet been achieved.

More recently the question of tackling BPS Wilson loops in more general representations has been successfully tackled at leading order. The introduction of general representations gives a new probing parameter, thus expanding the possibilities initiated in the context of the fundamental representation. In the holographic framework, a half BPS Wilson loop in $\mathcal{N}=4$ supersymmetric Yang-Mills (SYM) theory in the fundamental, symmetric or antisymmetric representation of $SU(N)$ is best described by a fundamental string, a D3-brane or a D5-brane with fluxes in their worldvolumes, respectively. Drukker and Fiol computed in \cite{Drukker:2005kx}, using a holographic D3 brane description, the expectation value of a $k$-winding circular string which, to leading order, coincides with the $k$-symmetric representation. A more rigorous analysis of the role of the representation was elucidated in \cite{Gomis:2006sb, Gomis:2006im}. Some progress on the questions of quantum corrections to these configurations immediately followed  with a strong emphasis on the field theory side  \cite{Yamaguchi2006, Hartnoll2006, Yamaguchi:2007ps}. Developing the gravity side of this correspondence is one of the main motivations for this work. In particular, we derive the spectrum of quantum fluctuations in the bosonic and fermionic sectors for a D5-brane with $k$ units of electric flux in its $AdS_2\times S^4$ world volume embedded in $AdS_5\times S^5$. This gravity configuration is the dual of the half BPS Wilson loop in the totally antisymmetric representation of rank $k$ in $\mathcal{N}=4$ SYM.

Although our main motivation comes from the study of Wilson loops, there is another strong motivation for our study of quantum fluctuations. String theory has heavily relied on the understanding of extended objects in the context of the gauge/gravity correspondence. They have played a key role in interpreting and identifying various hadronic configurations (quarks, baryons, mesons, $k$-strings). A more general approach on the quantization of these objects is a natural necessity. The long history of failed attempts at quantizing extended objects around flat space might have found its right context. Although largely motivated by holography, it is important by itself that the quantum theory of extended objects in asymptotically $AdS$ world volumes seems to be much better behaved than naively expected. In our simplified setup we are faced with various divergences, but many of them allow for some quite natural interpretations. Although we do not attack the general problem of divergences in a general context, we hope that our analysis could serve as a first step in this more fundamental direction of quantization of extended objects.

In this paper, we systematically study small fluctuations of D5-branes embedded in asymptotically $AdS_5\times S^5$, with flux in its world volume and wrapping an $S^4 \subset S^5$ \cite{Pawelczyk:2000hy, Camino:2001at, Hartnoll:2006ib, Armoni:2006ux}. The formalism we develop readily applies to more general backgrounds than just the holographic Wilson loop, including holographic Wilson loop correlators \cite{Zarembo:1999bu, Olesen:2000ji} and related finite-temperature configurations \cite{Hartnoll:2006hr, Kachru:2009xf}. Using this general formalism, we obtain the spectrum of both bosonic and fermionic excitations of D5-branes dual to the half BPS circular Wilson loop. Our analysis is explicit by nature and falls nicely in the group theoretic framework put forward in \cite{faraggi:2011bb}. We also  compute the one-loop effective action using heat kernel techniques.

The paper is organized as follows. In section~\ref{sec:background}, we introduce the class of D5-brane configurations for which our analysis applies. For completeness, the bulk background geometries and the main features of the D5-brane background configurations are reviewed in sections~\ref{review:bulk.bg} and \ref{review:D5.bg}, respectively. Section~\ref{fluct} contains the general analysis of the bosonic and fermionic excitations of these D5-branes. The second-order actions for the bosonic and fermionic degrees of freedom are constructed in sections~\ref{bosons} and \ref{fermions}, respectively, and their classical field equations are analyzed in sections~\ref{bosons:eoms} and \ref{fermions:eoms}.
Sections~\ref{spectrum} and \ref{det} deal with the holographic Wilson loop. The spectrum of fluctuations is obtained in section~\ref{spectrum}. Section~\ref{det} presents the calculation of the one-loop effective action using the heat kernel method. We conclude in section~\ref{sec:conclusions}. Technical material pertaining to our notation, the geometry of embeddings and to aspects of the heat kernel method are relegated to a series of appendices.

Note: While our work was in progress, the paper \cite{Harrison:2011fs} appeared, in which the spectrum of the bosonic fluctuations was derived.

%% file: Review_Bulk.tex
\section{Background geometry and classical D5-brane solutions}
\label{sec:background}
We begin by briefly reviewing the bulk geometry and classical D5-brane configurations we are interested in. Although we will eventually focus on $AdS_5\times S^5$, we emphasize that the methods developed in this paper are more general and apply to other solutions of type-IIB supergravity, including the near horizon limit of black D3-branes. Throughout the paper we will work in Lorentzian signature and switch to Euclidean signature only to discuss functional determinants in Sec.~\ref{det}. We refer the reader to Appendix \ref{convs} for notation and conventions.

\subsection{Bulk background}
\label{review:bulk.bg}

We are interested in probe D5-branes embedded in the following solution of type-IIB supergravity,
\begin{equation}
\label{review:bgsol}
	\begin{split}
	\rmd s^2 &= - f(r)\rmd t^2 +\frac{dr^2}{f(r)} +\frac{r^2}{\lad^2}\sum_{i=1}^3(\rmd x_i)^2
	+\lad^2 (\rmd\vt^2 +\sin^2\vt \rmd\Omega_4^2)~,\\
	C_{(4)} &=\frac{r^4}{\lad^4} \rmd t\wedge \rmd^3x +\lad^4 C(\vt) \, \rmd^4\Omega~,\\
	\end{split}
\end{equation}
where
\begin{equation}
\label{review:C.f}
	f(r)=\frac{r^2}{\lad^2}\left(1-\frac{\rh^4}{r^4}\right)~,\qquad
	C(\vt)= \frac32 \vt -\frac32 \sin\vt\cos\vt -\sin^3\vt\cos\vt~.
\end{equation}
All the other background fields vanish. The function $C(\vt)$ satisfies $\rmd C/\rmd \vt= 4 \sin^4 \vt$, so that the 5-form is
\begin{equation}
\label{review:F5}
	F_{(5)} = \rmd C_{(4)} = \frac{4}{\lad} (1+\ast) \rmd^5\Omega_\lad~,
\end{equation}
where $\rmd^5\Omega_\lad = \lad^5 \rmd^5 \Omega$ is the volume measure of a 5-sphere of radius $\lad$, as it appears in the metric \eqref{review:bgsol}.

With
\begin{equation}
\label{review:lad}
	\lad^4\equiv4\pi g_s N \alpha'^2 = \lambda \alpha'^2~,
\end{equation}
where $\lambda$ is the 't~Hooft coupling, \eqref{review:bgsol} describes $N$ D3-branes, generically at finite temperature. The black hole horizon ``radius'', $\rh$, is related to the inverse temperature by
\begin{equation}
\label{review:r.bh}
	\rh = \frac{\pi \lad^2}{\beta}~.
\end{equation}
The zero temperature $AdS_5 \times S^5$ solution is recovered by setting $\rh=0$. In this case, we can make the replacement $r\rightarrow L^2/z$ to obtain the $AdS_5$ metric in the standard Poincar\'e coordinates with boundary at $z=0$. Anticipating the embedding of the D5-branes, the metric of $S^5$ has been written in terms of an $S^4$ at some azimuth angle, $\vt \in [0,\pi]$. 

%% file: Review_D5.tex
\subsection{Background D5-branes}
\label{review:D5.bg}

In the background \eqref{review:bgsol}, the bosonic part of the D5-brane action is
\begin{equation}
\label{review:D5action.bos}
	S^{(B)}_{D5}
	     = - T_5 \int \rmd^6 \xi \sqrt{-\det(g+\F)_{ab}} + T_5 \int C_{(4)}\wedge \F~,
\end{equation}
where $T_5^{-1}= (2\pi)^5\alpha'{}^3 g_s$ is the D-brane tension, $g_{ab}$ is the induced metric on the world volume, and $\F=\rmd \A$ is the field strength living on the brane.

We consider D5-brane configurations such that four coordinates $\xi^{\mu}$ wrap the $S^4\subset S^5$ at a constant angle $\vt$ and the remaining two coordinates $\xi^{\alpha}=(\tau,\rho)$ span an effective string world sheet, with induced metric $g_{\alpha\beta}$, in the a$AdS_5$ part of the bulk.
By symmetry, the only non-vanishing components of $\F$ are\footnote{With a slight abuse of notation we use $\F$ to denote also the antisymmetric component.}
\begin{equation}
\label{review:F.def}
	\F_{\alpha\beta} = \F \epsilon_{\alpha\beta}~,
\end{equation}
and we can fix a gauge such that only $\A_\tau$ is non-trivial. It follows that
\begin{equation}
\label{review:det.gF}
	-\det(g+\F)_{ab} = -\det(g+\F)_{\alpha\beta}\det g_{\mu\nu} = \lad^8\sin^8 \vt\, (1-\F^2) (-\det g_{\alpha\beta})~.
\end{equation}
Hence, the action \eqref{review:D5action.bos} can be written as
\begin{equation}
\label{review:D5action.bos2}
	S^{(B)}_{D5}
	     = - \prefac \int \rmd\tau\rmd\rho \sqrt{-\det g_{\alpha\beta}} \left[ \sin^4\vt \sqrt{1-\F^2} - C(\vt) \F \right]~.
\end{equation}
The prefactor arises from $T_5 V_4\lad^4=\prefac$, where $V_4=8\pi^2/3$ is the volume of the unit $S^4$.

Quantization of 2-form flux (which is an integral of the equation of motion for $\A_\tau$), and the equation of motion for $\vt$ are solved by \cite{Pawelczyk:2000hy, Camino:2001at}
\begin{equation}
\label{review:n.theta}
	\nu \equiv \frac{n}{N} = \frac1\pi \left( \vt -\sin\vt\cos\vt \right)~,\qquad (n = 0,\ldots,N)~,	
\end{equation}
and
\begin{equation}
\label{review:f.sol}
	\F = \cos \vt~.
\end{equation}
Although $n$, the fundamental string charge dissolved on the D5-brane, is quantized, we can consider $\nu\in(0,1)$ as a continuous variable in the large-$N$ limit.

One must add to \eqref{review:D5action.bos2} appropriate boundary terms \cite{Drukker:1999zq, Drukker:2005kx}
\begin{equation}
\label{review:action.bt}
	\Sb^{(B)} = - \int \rmd \tau \sgn r'\, (r \pi_r +\A_\tau \pi_A)~,
\end{equation}
where
\begin{equation}
\label{review:momenta}
	\pi_r = \frac{\partial \Lag_{D5}}{\partial r'}~,\qquad
	\pi_A = \frac{\partial \Lag_{D5}}{\partial \A_\tau'} = -\frac1{\sqrt{-\det g_{\alpha\beta}}} \frac{\partial \Lag_{D5}}{\partial \F}~,
\end{equation}
and the prime denotes a derivative with respect to $\rho$.
Putting everything together, one finds that the action of the background D5-brane can be reduced to that of an effective string living in the a$AdS_5$
portion of the 10-dimensional geometry \cite{Hartnoll:2006ib}
\begin{equation}
\label{review:D5action.bos3}
	S^{(B)}_{D5} +\Sb^{(B)}
	     = - \prefac \sin^3\vt \left[ \int \rmd\tau\rmd\rho \sqrt{-\det g_{\alpha\beta}}
	     	- \int \rmd \tau \sgn r'\, r \frac{\partial \sqrt{-\det g_{\alpha\beta}}}{\partial r'} \right]~.
\end{equation} 

%% file: Fluctuations_Bosons.tex
\section{Fluctuations}
\label{fluct}

In this section, we consider the fluctuations of the bosonic and fermionic degrees of the D5-brane solutions described above. We construct the quadratic actions and derive the classical field equations. As a first result, the spectrum of fluctuations of the circular Wilson loop of operators in the anti-symmetric representations, predicted in \cite{faraggi:2011bb}, is fully derived. Our formalism readily applies to more general backgrounds, including holographic Wilson loop correlators \cite{Zarembo:1999bu, Olesen:2000ji} and related finite-temperature configurations \cite{Hartnoll:2006hr, Kachru:2009xf}.

\subsection{Bosonic fluctuations}
\label{bosons}
Let us start by defining the dynamical variables that parameterize the physical fluctuations. We will make use of well-known geometric relations for embedded manifolds \cite{Eisenhart}, which are reviewed in appendix~\ref{embed}. The fields present in \eqref{review:D5action.bos} are the target-space coordinates of the D5-brane $x^m$ and the gauge field components $\mathcal{A}_a$ living on the brane. Both are functions of the D5-brane world volume coordinates $\xi^a$.

We now recall a few facts from differential geometry that, although known to the reader, we bring to bear explicitly in our calculations. We shall parameterize the fluctuations of $x^m$ around the background coordinates by the generating vector $y$ of an exponential map \cite{Eisenhart}
\begin{equation}
\label{bosons:geo.exp}
	x^m \to (\exp_x y)^m = x^m +y^m -\frac12 \Gamma^m{}_{np} y^n y^p +\mathcal{O}(y^3)~,
\end{equation}
thereby obtaining a formulation that is manifestly invariant under bulk diffeomorphisms. Recall that, as familiar from General Relativity, the differences of coordinates are not covariant objects, but vector components are. Here and henceforth, all quantities except the fluctuation variables are evaluated on the background. Locally, the vector components $y^m$ coincide with the Riemann normal coordinates centered at the origin of the exponential map. Riemann normal coordinates are also helpful for performing the calculations, because of a number of simplifying relations that hold at the origin. For example, one can make use of
\begin{equation}
\label{bosons:riem.coord}
	\Gamma^m{}_{np}=0~,\qquad \Gamma^m{}_{np,q} = -\frac23 R^m{}_{(np)q}~,
\end{equation}
while the expression for a covariant tensor of rank $k$ is, up to second order in $y$,
\begin{equation}
\label{bosons:tensor.expand}
	A_{m_1\ldots m_k} \to A_{m_1\ldots m_k} + A_{m_1\ldots m_k;n} y^n + \frac12\left( A_{m_1\ldots m_k;np}
	+\frac13 \sum\limits_{l=1}^k R^q{}_{npm_l} A_{m_1\ldots q\ldots m_k} \right) y^n y^p~.
\end{equation}
In the equations that follow, we will implicitly assume the use of a Riemann normal coordinate system. Moreover, we shall drop terms of higher than second order in $y$.  The tangent vectors along the world volume (see appendix~\ref{embed}), which serve to calculate the pull-back of bulk tensor fields, are given by
\begin{equation}
\label{bosons:tangent}
	x^m_a \to x^m_a +\nabla_a y^m -\frac13 R^m{}_{pnq} x^n_a y^p y^q~.
\end{equation}
Reparametrization invariance allows us to gauge away the fluctuations that are tangent to the world volume. This leaves us with
\begin{equation}
\label{bosons:fluct.normal}
	y^m = N^m_\flatind{i} \chi^\flatind{i}~,
\end{equation}
where the $\chi^\flatind{i}$ parameterize the fluctuations orthogonal to the world volume, or normal fluctuations, and the index $\flatind{i}$ runs over all normal directions. The expression above is the natural geometric object related to fluctuations; it has appeared in previous works, for example, \cite{martucci:2005rb} and, more explicitly, in \cite{faraggi:2011bb}. We found it appropriate to provide an explicit account of the origin of this parametrization of the fluctuations. Using the relations summarized in appendix~\ref{embed}, this gives rise to
\begin{equation}
\label{bosons:fluct.tangent}
	\nabla_a y^m = -H_{\flatind{i}a}{}^b x^m_b \chi^\flatind{i} + N^m_\flatind{i} \nabla_a \chi^\flatind{i}~,
\end{equation}
where $H_{\flatind{i}a}{}^b$ is the second fundamental form of the background world volume, and $\nabla_a$ denotes the covariant derivative including the connections in the normal bundle.

The fluctuations of the gauge field are introduced by
\begin{equation}
\label{bosons:fluct.F}
	\F_{ab} \to \F_{ab} + f_{ab}~,
\end{equation}
and we recall that the background gauge field only lives on the 2-d part of the world volume as shown in \eqref{review:F.def}.

Following these preliminaries, we now consider fluctuations of the degrees of freedom of the D5-brane. The goal is to expand the action \eqref{review:D5action.bos} to second order in the fields $\chi^{\underline{i}}$ and $a_a$. For the Born-Infeld term, we make use of the formula
\begin{equation}
\label{bosons:BI.expand}
	 \sqrt{-\det M} \to \sqrt{-\det M}\left\{ 1 + \frac12 \Tr X +\frac18 [\Tr X]^2
	 -\frac14 \Tr(X X) +\mathcal{O}(X^3) \right\}~,
\end{equation}
where $X$ denotes the matrix $X=M^{-1}\delta M$, and we have introduced
\begin{equation}
\label{bosons:M.def}
	M_{ab} = g_{ab} +\F_{ab}~.
\end{equation}
Combining \eqref{bosons:tensor.expand}--\eqref{bosons:fluct.tangent} to obtain the induced metric, we have
\begin{equation}
\label{bosons:delta.M}
	\delta M_{ab} = -2H_{\flatind{i}ab} \chi^\flatind{i} + f_{ab}
	+ \nabla_a \chi^\flatind{i} \,\nabla_b \chi^\flatind{j} \,\delta_\flatind{ij}
	+  \left( H_{\flatind{i}a}{}^c H_{\flatind{j}bc} -R_{mpnq} x^m_a x^n_b N^p_\flatind{i} N^q_\flatind{j} \right)
	\chi^\flatind{i} \chi^\flatind{j}~.
\end{equation}

Substituting \eqref{bosons:delta.M} into \eqref{bosons:BI.expand} and using the background relations, one obtains after some calculation
\begin{equation}
\label{bosons:BI.action}
\begin{split}
	\sqrt{-\det M_{ab}} &\to \lad^4 \sin^3 \vt \sqrt{-\det g_{\alpha\beta}} \left[ \sin^2\vt+\frac4{\lad} \cos\vt\sin\vt\, \chi^\flatind{5} +
	\frac12\cos\vt\,\epsilon^{\alpha\beta} f_{\alpha\beta}
	+\frac{2\cos^2\vt}{\lad\sin\vt} \chi^\flatind{5} \epsilon^{\alpha\beta} f_{\alpha\beta}  \right.\\
	&\quad +\frac12 \hat{g}^{ab} \left(\delta_\flatind{ij} \nabla_a \chi^\flatind{i} \nabla_b \chi^\flatind{j}
	+\nabla_a \chi^\flatind{5} \nabla_b \chi^\flatind{5} \right) +\frac1{4\sin^2\vt} \hat{g}^{ab} \hat{g}^{cd} f_{ac} f_{bd} \\
	&\quad \left. +\frac2{\lad^2}\left(3 \cos^2\vt-\sin^2\vt\right) (\chi^\flatind{5})^2
	-\frac12 \left( H_{\flatind{i}\alpha\beta}H_\flatind{j}{}^{\alpha\beta} +
	R_{mpnq}\, g^{\alpha\beta} x_\alpha^m x_\beta^n N^p_\flatind{i} N^q_\flatind{j} \right) \chi^\flatind{i} \chi^\flatind{j} \right]~,
\end{split}
\end{equation}
where $\hat{g}^{ab}$ is the inverse of the 6-d metric
\begin{equation}
\label{bosons:hat.g}
	\rmd\hat{s}^2 = g_{\alpha \beta} \rmd \xi^\alpha \rmd \xi^\beta + \lad^2 \rmd\Omega_4^2~
\end{equation}
which is independent of $\vt$. Henceforth, the normal index $\flatind{i}$ in \eqref{bosons:BI.action} refers only to the three normal directions within the $AdS_5$ part of the bulk, as we have indicated explicitly $\chi^\flatind{5}$ for the normal direction within $S^5$. Note that \eqref{bosons:hat.g} is not the induced metric on the worldvolume. Rather, the background flux deforms the metric and the fluctuations see the deformed geometry as appropriate for open string fluctuations.

In order to expand the Chern-Simons term in \eqref{review:D5action.bos}, we make use of \eqref{bosons:tensor.expand}, \eqref{bosons:tangent}, \eqref{bosons:fluct.F} and the background relations. One soon finds that to quadratic order in the fluctuations, only the components of the four-form $C_{(4)}$ that live on $S^4$ contribute. After some calculation one obtains
\begin{equation}
\label{bosons:CS.action}
\begin{split}
	C\wedge \mathcal{F} &\to \rmd^4\Omega \rmd \tau \rmd\rho \sqrt{-\det g_{\alpha\beta}} \lad^4\left[ C(\vt) \cos\vt
	-\frac12 C(\vt) \epsilon^{\alpha\beta} f_{\alpha\beta} +\frac4{\lad} \sin^4\vt \cos\vt\, \chi^\flatind{5} \right.\\
	&\quad \left.
	-\frac2{\lad} \sin^4\vt \,\chi^\flatind{5} \epsilon^{\alpha\beta} f_{\alpha\beta} +\frac8{\lad^2} \sin^3\vt\cos^2\vt\,
	(\chi^\flatind{5})^2 \right]~.
\end{split}
\end{equation}

Replaceing \eqref{bosons:BI.action} and \eqref{bosons:CS.action} in \eqref{review:D5action.bos}, the linear terms in $\chi^\flatind{5}$ are found to cancel as expected for an expansion around a classical solution. The linear term in $f_{\alpha\beta}$ is a total derivative and is cancelled by a boundary term similar to \eqref{review:action.bt}.
Thus, one ends up with the following quadratic terms in the action,
\begin{equation}
\label{bosons:action}
\begin{split}
	S^{(B,2)}_{D5} &= T_5 \sin^3\vt \int \rmd^6 \xi \sqrt{-\det \hat{g}_{ab}}
	 \left[ -\frac12 \hat{g}^{ab} \left(\delta_\flatind{ij} \nabla_a \chi^\flatind{i} \nabla_b \chi^\flatind{j}
	+\nabla_a \chi^\flatind{5} \nabla_b \chi^\flatind{5} \right) +\frac2{\lad^2} (\chi^\flatind{5})^2 \right.\\
	&\quad \left.
	-\frac1{4\sin^2\vt} \hat{g}^{ab} \hat{g}^{cd} f_{ac} f_{bd}
	-\frac{2}{\lad\sin\vt} \chi^\flatind{5} \epsilon^{\alpha\beta} f_{\alpha\beta}
	+\frac12 \left( H_{\flatind{i}\alpha\beta}H_\flatind{j}{}^{\alpha\beta}
	+ R_{mpnq}\, g^{\alpha\beta} x_\alpha^m x_\beta^n N^p_\flatind{i} N^q_\flatind{j} \right) \chi^\flatind{i} \chi^\flatind{j} \right]~.
\end{split}
\end{equation}
The dynamical fields present in \eqref{bosons:action} are the scalar $\chi^\flatind{5}$, the scalars $\chi^\flatind{i}$ transforming as a triplet under the $SO(3)$ symmetry of the normal bundle, and the gauge fields $a_a$, $f_{ab}=\partial_a a_b-\partial_b a_a$.

%% file: Fluctuations_Fermions.tex
\subsection{Fermionic fluctuations}
\label{fermions}

We now consider fluctuations of the fermionic degrees of freedom of the D5-brane. This is somewhat easier than the bosonic part, because one just needs the fermionic part of the action, in which all bosonic fields assume their background values.

Our starting point is the fermionic part of the D5-brane action with $\kappa$-symmetry, which was derived in \cite{martucci:2005rb}\footnote{We thank L.~Martucci for pointing out to us that, in order to correctly  interpret the symbol $W_m$ in the gauge-fixed action (30) of \cite{martucci:2005rb}, one should start from their equation (17). We have omitted the term $\Delta$, which vanishes in the background \eqref{review:bgsol}.}
\begin{equation}
\label{fermions:action}
	S^{(F)}_{D5}
	     = \frac{T_5}2 \int \rmd^6 \xi \sqrt{-\det M_{ab}}\, \bar{\theta} \left(1-\Gamma_{D5} \right)
	     \left[ (\tilde{M}^{-1})^{ab} \Gamma_b D_a \right] \theta~,
\end{equation}
where $M_{ab}=(g+\F)_{ab}$ as before, $\tilde{M}_{ab} = g_{ab} + \sigma_3 \Gamma_{(10)} \F_{ab}$, $\theta$ is a doublet of 32-component left-handed Majorana-Weyl spinors ($\Gamma_{(10)}\theta = \theta$, the $\sigma_3$ acts on the doublet notation) and $\bar{\theta}=i \theta^\dagger\Gamma^\flatind{0}$. Moreover, in the background \eqref{review:bgsol}, the derivative operator $D_a$ is given by
\begin{equation}
\label{fermions:Da}
	D_a = \left( \partial_a x^m\right) \left[ \nabla_m + \frac{i\sigma_2}{16\cdot 5!} F_{npqrs}\Gamma^{npqrs} \Gamma_m \right]~,
\end{equation}
where $\nabla_m=\partial_m +\frac14 \omega_m^{\;\;\,\flatind{np}} \Gamma_\flatind{np}$ is the 10-d spinor covariant derivative. We give a derivation of its pull-back onto the world volume in appendix~\ref{embed}. Using \eqref{embed:D} and the background relations \eqref{review:bgsol}, one finds
\begin{equation}
\label{fermions:Dalpha}
	D_\alpha = \nabla_\alpha +\frac14 A_{\flatind{ij} \alpha} \Gamma^\flatind{ij}
	-\frac12 H_{\flatind{i}\alpha\beta} \Gamma^\beta \Gamma^\flatind{i}
	- \frac1{4\lad} \Gamma_\alpha \Gamma^\flatind{5\cdots9} (i\sigma_2) \left( 1+\Gamma_{(10)} \right)
\end{equation}
and
\begin{equation}
\label{fermions:Dmu}
	D_\mu = \nabla_\mu -\frac12 H_{\flatind{i}\mu\nu} \Gamma^\nu \Gamma^\flatind{5}
	+ \frac1{4\lad} \Gamma_\mu \Gamma^\flatind{5\cdots9} (i\sigma_2) \left( 1+\Gamma_{(10)} \right)~.
\end{equation}
The matrix $\Gamma_{D5}$ in \eqref{fermions:action} is\footnote{The matrix $\Gamma^{(0)}_{D5}$ of \cite{martucci:2005rb} is given by $\Gamma^{(0)}_{D5} = -\tilde{\Gamma} \Gamma^\flatind{6\cdots9}$. Notice the minus sign due to their definition of the epsilon symbol, which differs from ours.}
\begin{equation}
\label{fermions:GammaD5}
	\Gamma_{D5} = \frac1{\sin\vt} \tilde{\Gamma} \Gamma^\flatind{6\cdots9}\, \sigma_1
	\left( 1+\sigma_3 \tilde \Gamma \cos \vt\right)~,
\end{equation}
where we have defined
\begin{equation}
\label{fermions:tilde.Gamma}
	\tilde{\Gamma} = \frac12 \epsilon_{\alpha\beta} \Gamma^{\alpha\beta}~.
\end{equation}
It is useful to re-write \eqref{fermions:GammaD5} as
\begin{equation}
\label{fermions:GammaD5.2}
	\Gamma_{D5} = \e{R\sigma_3\tilde{\Gamma}} \tilde{\Gamma} \Gamma^\flatind{6\cdots9} \sigma_1
	\e{-R\sigma_3\tilde{\Gamma}}~, \qquad \sinh(2R) =-\cot\vt~.
\end{equation}
Finally, the inverse of the matrix $\tilde{M}_{ab}$ is found to be
\begin{equation}
\label{fermions:Ms}
	\left(\tilde{M}^{-1}\right)^{\mu\nu} = g^{\mu\nu}~,\qquad
	\left(\tilde{M}^{-1}\right)^{\alpha\beta} = \frac1{\sin^2\vt}
		\left( g^{\alpha\beta} - \cos \vt \,\epsilon^{\alpha\beta} \sigma_{3} \Gamma_{(10)} \right)~.\end{equation}

Now, owing to the fact that $\theta$ is left-handed, we can replace $\Gamma_{(10)}$ by $+1$ in \eqref{fermions:Dalpha} and \eqref{fermions:Dmu} and by $-1$ in \eqref{fermions:Ms}, when they are substituted into \eqref{fermions:action}. Because $\epsilon^{\alpha\beta}\Gamma_\beta = -\tilde{\Gamma}\Gamma^\alpha$, we also find
\begin{equation}
\label{fermions:M.gamma}
	\left(\tilde{M}^{-1}\right)^{\alpha\beta} \Gamma_\beta = \frac1{\sin\vt}
		\e{R\sigma_3\tilde{\Gamma}} \Gamma^\alpha \e{-R\sigma_3 \tilde{\Gamma}}
\end{equation}
when acting on a left-handed spinor. Putting everything together and using also that the extrinsic curvature terms in \eqref{fermions:Dalpha} and \eqref{fermions:Dmu} are $H^{\flatind{5}\mu}{}_\nu = -(\cot\vt/\lad) \delta^\mu_\nu$ and $H_{\flatind{i}\alpha}{}^\alpha=0$, because the 2-d part of the background is a minimal surface, we find after some calculation that the action \eqref{fermions:action} becomes
\begin{equation}
\label{fermions:action2}
	S^{(F)}_{D5} = \frac{T_5}2 \sin^4\vt \int \rmd^6 \xi \sqrt{-\det \hat{g}_{ab}}\, \bar{\theta}
	\e{R\sigma_3\tilde{\Gamma}} \left( 1- \tilde\Gamma\Gamma^\flatind{6\cdots9} \sigma_1 \right)
	\left[ \hat{\Gamma}^\mu \nabla_\mu +\Gamma^\alpha \tilde{\nabla}_\alpha
	+\frac1{\lad} \Gamma^\flatind{5\cdots9} (i\sigma_2) \right] \e{-R\sigma_3\tilde{\Gamma}} \theta~,
\end{equation}
where we have abbreviated
\begin{equation}
\label{fermions:tilde.nabla}
	\tilde{\nabla}_\alpha = \nabla_\alpha +\frac14 A_{\flatind{ij}\alpha} \Gamma^\flatind{ij}
\end{equation}
to denote the covariant spinor derivative including the connections in the normal bundle, $\hat{\Gamma}^\mu=\sin\vt \Gamma^\mu$ are the covariant gamma matrices normalized for a 4-sphere of radius $\lad$, and $\hat{g}_{ab}$ is the 6-d metric \eqref{bosons:hat.g}, which we used also for the bosons.

To simplify \eqref{fermions:action2} slightly, we introduce the rotated double spinor $\theta'= \e{-R\sigma_3 \tilde{\Gamma}} \theta$. Its conjugate is easily found as $\bar{\theta'} = i \theta^\dagger \e{-R\sigma_3 \tilde{\Gamma}} \Gamma^\flatind{0} = \bar{\theta} \e{R\sigma_3 \tilde{\Gamma}}$, which is just the combination that appears in \eqref{fermions:action2}. Henceforth, we shall work with the rotated spinor and drop the prime for brevity.

Now we fix the $\kappa$-symmetry. The covariant gauge-fixing condition $\Gamma_{(10)} \sigma_3 \theta= \theta$ \cite{martucci:2005rb} reduces to $\sigma_3 \theta = \theta$, because $\theta$ is left-handed. The terms in the action that survive this projection are
\begin{equation}
\label{fermions:action3}
	S^{(F)}_{D5} = \frac{T_5}2 \sin^4\vt \int \rmd^6 \xi \sqrt{-\det \hat{g}_{ab}}\, \bar{\theta}
	\left[ \hat{\Gamma}^\mu \nabla_\mu +\Gamma^\alpha \tilde{\nabla}_\alpha
	+\frac1{\lad} \tilde{\Gamma} \Gamma^\flatind{5} \right] \theta~,
\end{equation}
where $\theta$ is now a single, 32-component spinor. 

Next, let us write \eqref{fermions:action3} in terms of 6-d spinors. For this purpose, we choose the following, chiral representation of the 10-d gamma matrices,
\begin{equation}
\label{fermions:gamma.ten.decomp}
\begin{aligned}
	\Gamma^\flatind{\alpha} &= \Ifour \otimes \gamma^\flatind{\alpha}
		\otimes \Itwo \otimes \sigma_2 \qquad \qquad& (\alpha&=0,1) \\
	\Gamma^\flatind{i} &= \Ifour \otimes \gamma^\flatind{01}
		\otimes \tau^\flatind{i} \otimes \sigma_2 & (i&= 2,3,4) \\
	\Gamma^\flatind{5} &= \hat{\gamma}^\flatind{5} \otimes \Itwo
		\otimes \Itwo \otimes \sigma_1   \\
	\Gamma^\flatind{\mu} &= i\hat{\gamma}^\flatind{5} \hat{\gamma}^\flatind{\mu} \otimes \Itwo
		\otimes \Itwo \otimes \sigma_1 & (\mu&=6,7,8,9)~,
\end{aligned}
\end{equation}
where $\hat{\gamma}^\flatind{\mu}$, $\gamma^\flatind{\alpha}$ and $\tau^\flatind{i}$ are Euclidean 4-d gamma matrices, Lorentzian 2-d gamma matrices and a set of Pauli matrices, respectively, satisfying
\begin{equation}
\label{fermions:gamma.clifford}
	\left\{ \hat{\gamma}^\flatind{\mu}, \hat{\gamma}^\flatind{\nu} \right\}
		= 2\delta^\flatind{\mu\nu}~, \qquad
	\left\{ \gamma^\flatind{\alpha}, \gamma^\flatind{\beta} \right\}
		= 2\eta^\flatind{\alpha\beta}~, \qquad
	\left\{ \tau^\flatind{i}, \tau^\flatind{j} \right\} = 2\delta^\flatind{ij}~,
\end{equation}
and $\hat{\gamma}^\flatind{5} = \hat{\gamma}^\flatind{6789}$. Notice the peculiar representation of $\Gamma^\flatind{\mu}$, which will turn out to be handy for reconstructing 6-d gamma matrices.
It follows from \eqref{fermions:gamma.ten.decomp} that the 10-d chirality matrix is simply
\begin{equation}
\label{fermions:gamma.ten.chiral}
	\Gamma_{(10)} = \Gamma^\flatind{0\cdots9} = \Ifour \otimes \Itwo \otimes \Itwo \otimes \sigma_3~.
\end{equation}

Then, after reconstructing the 6-d gamma matrices for the D5-brane world volume by
\begin{equation}
\label{fermions:six.decomp}
	\hat{\Gamma}^\flatind{\alpha} = \hat{\gamma}^\flatind{5} \otimes \gamma^\flatind{\alpha}~,
	\qquad
	\hat{\Gamma}^\flatind{\mu} = \hat{\gamma}^\flatind{\mu} \otimes \Itwo~,
\end{equation}
and using the left-handedness of $\theta$, the action \eqref{fermions:action3} becomes
\begin{equation}
\label{fermions:action4}
	S^{(F)}_{D5} = \frac{T_5}2 \sin^4\vt \int \rmd^6 \xi \sqrt{-\det \hat{g}_{ab}}\, \bar{\theta}
	\left[ \hat{\Gamma}^a \nabla_a
	+\frac14 \hat{\Gamma}^\alpha A_{\flatind{ij}\alpha} \tau^\flatind{ij}
	-\frac{i}{\lad} \hat{\Gamma}^\flatind{01} \right] \theta~,
\end{equation}
where $\theta$ now represents a doublet of 8-component, 6-d Dirac spinors that stems from the spinor components that survive the chiral projection. The 6-d gamma matrices $\hat{\Gamma}^a$ act on the spinors, and the matrices $\tau^\flatind{ij}=i\epsilon^\flatind{ijk} \tau^\flatind{k}$ act on the doublet.

Performing a chiral rotation, $\theta= \e{i \eta \hat{\Gamma}_{(6)}} \theta'$, one can obtain other, equivalent ways of writing the action \eqref{fermions:action4}, in which the ``mass'' term changes its appearance. In particular, for $\eta=\pi/4$, one obtains
\begin{equation}
\label{fermions:action5}
	S^{(F)}_{D5} = \frac{T_5}2 \sin^4\vt \int \rmd^6 \xi \sqrt{-\det \hat{g}_{ab}}\, \bar{\theta}'
	\left[ \hat{\Gamma}^a \nabla_a
	+\frac14 \hat{\Gamma}^\alpha A_{\flatind{ij}\alpha} \tau^\flatind{ij}
	+\frac{1}{\lad} \hat{\Gamma}^\flatind{6789} \right] \theta'~.
\end{equation}
In contrast to \eqref{fermions:action4}, in which the mass term commutes with the 4-d part of the kinetic term and anti-commutes with the 2-d part, in \eqref{fermions:action5} it commutes with the 2-d part of the kinetic term and anti-commutes with the 4-d part. In Sec.~\ref{det}, \eqref{fermions:action4} and \eqref{fermions:action5} will give rise to two different ways of calculating the heat kernel, with slightly different results.

To conclude the 6-d reduction of the fermionic action, we consider the 10-d Majorana condition.
In the decomposition \eqref{fermions:gamma.ten.decomp}, the intertwiner is given by\footnote{It is irrelevant whether we use $B_{+(9,1)}$ or $B_{-(9,1)}$, because they differ by a factor of $\Gamma_{(10)}$.}
\begin{equation}
\label{fermions:B.ten}
	B_{+(9,1)} = B_{+(5,0)} \otimes B_{-(1,1)} \otimes B_{-(3,0)} \otimes \Itwo~.
\end{equation}
Notice that the choice of intertwiners is unique for the odd-dimensional parts, but we have indicated the signs for clarity. Moreover, from \eqref{fermions:gamma.ten.decomp} it is evident that $B_{+(5,0)}= B_{-(4,0)}$ when acting on the $\hat{\gamma}^\mu$. Therefore, the first two factors on the right hand side of \eqref{fermions:B.ten} just form the 6-d intertwiner
\begin{equation}
\label{fermions:B.six}
	B_{-(5,1)} = B_{-(4,0)} \otimes B_{-(1,1)}~.
\end{equation}
Thus, if we write the 6-d spinor doublet as
\begin{equation}
\label{fermions:spinor.doublet}
	\theta = \sum_{r=\pm1} \theta^r \otimes \eta_r~,
\end{equation}
where $\eta_r$ is a doublet of constant (and normalized) 3-d symplectic Majorana spinors satisfying
\begin{equation}
\label{fermions:majorana.three}
	\eta_r{}^\ast = r B_{-(3,0)} \eta_{-r}~,
\end{equation}
then the 10-d Majorana condition $\theta^\ast= B_{+(9,1)}\theta$ gives rise to the symplectic Majorana condition on the two 6-d spinors,
\begin{equation}
\label{fermions:majorana.six}
	\theta^r{}^\ast = r B_{-(5,1)} \theta^{-r}~.
\end{equation}
A similar analysis can be done for the chirally rotated spinor $\theta'$. This completes the 6-d formulation of the fermionic action.

%% file: Fluctuations_EOMbosons.tex
\subsection{Classical field equations - bosons}
\label{bosons:eoms}

We shall work in the Lorentz gauge
\begin{equation}
\label{bosons:gauge.cond}
	\hat{\nabla}_a a^a = 0~,
\end{equation}
where $\hat{\nabla}_a$ denotes the covariant derivative with respect to the metric \eqref{bosons:hat.g} and, if acting on fields with indices $\flatind{i}$, contains also the appropriate connections for the normal bundle.
Condition \eqref{bosons:gauge.cond} leaves the residual gauge symmetry $a_a \to a_a+\partial_a \lambda$ with $\hat{\nabla}^a \hat{\nabla}_a \lambda=0$. Taking this into account, the field equations that follow from \eqref{bosons:action} are
\begin{align}
\label{bosons:chi.i.eq}
	\left[ \delta^\flatind{i}_\flatind{j}\, \hat{\nabla}^a \hat{\nabla}_a
	+ H^\flatind{i}{}_{\alpha\beta} H_\flatind{j}{}^{\alpha\beta} +R_{mpnq}\,g^{\alpha \beta} x^m_\alpha x^n_\beta N^{\flatind{i}p} N_\flatind{j}^q \right]
	\chi^\flatind{j} &=0~,\\
\label{bosons:chi.five.eq}
	\left(\hat{\nabla}^a \hat{\nabla}_a +\frac4{\lad^2}\right) \chi^\flatind{5} - \frac4{\lad \sin\vt} \epsilon_{\alpha\beta} \nabla^\alpha a^\beta
	&=0~,\\
\label{bosons:a.alpha.eq}
	\left( \hat{\nabla}^a \hat{\nabla}_a -\frac12 R_{(2)} \right) a^\alpha -\frac{4\sin\vt}{\lad} \epsilon^{\alpha\beta}
	\nabla_\beta \chi^\flatind{5} &=0~,\\
\label{bosons:a.mu.eq}
	\left( \hat{\nabla}^a \hat{\nabla}_a -\frac{3}{\lad^2} \right) a^\mu &=0~.
\end{align}
Note that $\hat{\nabla}^a \hat{\nabla}_a= \hat{\nabla}^\mu \hat{\nabla}_\mu + \nabla^\alpha\nabla_\alpha$, where $\hat{\nabla}^\mu \hat{\nabla}_\mu$ is the Laplacian on an $S^4$ of radius $\lad$, while the covariant derivative $\nabla_\alpha$ on the 2-d part of the world volume includes the connections for the normal bundle in the case of \eqref{bosons:chi.i.eq}. In \eqref{bosons:a.alpha.eq}, $R_{(2)}$ denotes the curvature scalar of the 2-d part of the world volume. So far, the components $a_\mu$ and $a_\alpha$ of the gauge fields are not entirely decoupled from each other, because of the gauge condition \eqref{bosons:gauge.cond}. However, we can use the residual gauge freedom to set $\nabla_\alpha a^\alpha=0$ on-shell. To see this, contract \eqref{bosons:a.alpha.eq} with $\nabla_\alpha$, which yields
\begin{equation}
\label{bosons:da.alpha.eq}
	\hat{\nabla}^a \hat{\nabla}_a \nabla_\alpha a^\alpha =0~.
\end{equation}
Thus, for any $a^\alpha$ satisfying \eqref{bosons:da.alpha.eq}, one can find a residual gauge transformation $\lambda$ satisfying $\nabla^\alpha\nabla_\alpha \lambda +\nabla_\alpha a^\alpha=0$ making the fields $a_\mu$ and $a_\alpha$ transverse,
\begin{equation}
\label{bosons:gauge.cond2}
	\nabla_\alpha a^\alpha = \hat{\nabla}_\mu a^\mu=0~.
\end{equation}
This still leaves us with the residual gauge transformations satisfying
\begin{equation}
\label{bosons:res.gauge}
	\hat{\nabla}^\mu\hat{\nabla}_\mu \lambda = \nabla^\alpha\nabla_\alpha\lambda=0~.
\end{equation}

To continue, we decompose the fields into\footnote{Notice the index shift for the vector harmonics, which is used to have all sums start from $l=0$. The sums over other quantum numbers are implicit.}
\begin{equation}
\label{bosons:decomp}
\begin{aligned}
	\chi^\flatind{j} &= \sum_{l=0}^\infty \chi^\flatind{j}_l(\tau,\rho) Y_l(\Omega)~, \qquad &
	\chi^\flatind{5} &= \sum_{l=0}^\infty \chi^\flatind{5}_l(\tau,\rho) Y_l(\Omega)~,\\
	a^\alpha &= \sum_{l=0}^\infty a^\alpha_l(\tau,\rho) Y_l(\Omega)~, &
	a^\mu &= \sum_{l=0}^\infty a_l(\tau,\rho) Y_{l+1}^\mu(\Omega)~,
\end{aligned}
\end{equation}
where $Y_l(\Omega)$ and $Y_{l+1}^\mu(\Omega)$ are scalar and transverse vector eigenfunctions of the Laplacian on $S^4$, respectively. The corresponding eigenvalues and their degeneracies are given by \cite{rubin:2888}
\begin{align}
\label{bosons:ev.0}
	\hat{\nabla}^\mu \hat{\nabla}_\mu Y_l(\Omega) &= -\frac{l(l+3)}{\lad^2} Y_l(\Omega)~, &
	D_l(4,0) &= \frac16(l+1)(l+2)(2l+3)~,\\
\label{bosons:ev.1}
	\hat{\nabla}^\mu \hat{\nabla}_\mu Y^\nu_{l+1}(\Omega) &= -\frac{l^2 +5l+3}{\lad^2} Y^\nu_{l+1}(\Omega)~, &
	D_{l+1}(4,1) &= \frac12 (l+1)(l+4)(2l+5)~.
\end{align}

Substituting \eqref{bosons:decomp}, \eqref{bosons:ev.0} and \eqref{bosons:ev.1} into the field equations \eqref{bosons:chi.i.eq}--\eqref{bosons:a.mu.eq} yields
\begin{align}
\label{bosons:chi.i.eq2}
	\left[ \left(\nabla^\alpha \nabla_\alpha -\frac{l(l+3)}{\lad^2} \right) \delta^\flatind{i}_\flatind{j}
	+ H^\flatind{i}{}_{\alpha\beta} H_\flatind{j}{}^{\alpha\beta} +R_{mpnq}\, g^{\alpha \beta}x^m_\alpha x^n_\beta N^{\flatind{i}p} N_\flatind{j}^q \right]
	\chi^\flatind{j}_l &=0~,\\
\label{bosons:chi.five.eq2}
	\left(\nabla^\alpha \nabla_\alpha -\frac{l(l+3)-4}{\lad^2} \right) \chi^\flatind{5}_l
	- \frac4{\lad \sin\vt} \epsilon_{\alpha\beta} \nabla^\alpha a^\beta_l &=0~,\\
\label{bosons:a.alpha.eq2}
	\left( \nabla^\beta \nabla_\beta -\frac{l(l+3)}{\lad^2} -\frac12 R_{(2)} \right) a^\alpha_l
	-\frac{4\sin\vt}{\lad} \epsilon^{\alpha\beta} \nabla_\beta \chi^\flatind{5}_l &=0~,\\
\label{bosons:a.mu.eq2}
	\left( \nabla^\alpha \nabla_\alpha -\frac{(l+2)(l+3)}{\lad^2} \right) a_l &=0~.
\end{align}

The dynamics of the two components $a^\alpha$ is contained in the field strength $f= \epsilon^{\alpha\beta} \nabla_\alpha a_\beta$. After decomposing $f$ into spherical harmonics on $S^4$, one can proceed to diagonalize \eqref{bosons:chi.five.eq2} and \eqref{bosons:a.alpha.eq2}, which gives rise to the 2-d Klein-Gordon equations
\begin{align}
\label{bosons:chi.a.diag1}
	\left[ \nabla^\alpha \nabla_\alpha - \frac1{\lad^2} (l+3)(l+4) \right] \zeta_l&=0~,&
	\zeta_l &= \left[ f_l +\frac{\sin\vt}{\lad}(l-1) \chi^\flatind{5}_l \right]~,\\
\label{bosons:chi.a.diag2}
	\left[ \nabla^\alpha \nabla_\alpha - \frac1{\lad^2} l(l-1) \right]\eta_l&=0 ~, &
	\eta_l &= \left[ f_l -\frac{\sin\vt}{\lad}(l+4) \chi^\flatind{5}_l \right]~.
\end{align}
We should exclude the $l=0$ case of \eqref{bosons:chi.a.diag2}, because in this case one can rewrite \eqref{bosons:a.alpha.eq2} identically as
\begin{equation}
\label{bosons:peculiar.mode}
	\epsilon^{\alpha\beta} \nabla_\beta \eta_0=0~,
\end{equation}
which implies that this particular mode is not dynamical. A similar result was found in \cite{Camino:2001at}. This matches with the fact that the residual gauge transformation \eqref{bosons:res.gauge} is given by a 2-d massless field with $SO(5)$ angular momentum $l=0$.

To summarize, the classical field equations for the bosonic fluctuations have been reduced to the 2-d field equations \eqref{bosons:chi.i.eq2}, \eqref{bosons:a.mu.eq2}, \eqref{bosons:chi.a.diag1} and \eqref{bosons:chi.a.diag2}. 

%% file: Fluctuations_EOMfermions.tex
\subsection{Classical field equations - fermions}
\label{fermions:eoms}

Let us consider the classical field equations for the fermions. We shall be agnostic about the symplectic Majorana condition \eqref{fermions:majorana.six}, which can be imposed afterwards. This has the advantage that the following arguments hold also if we switch to Euclidean signature. The Dirac equation following from the action \eqref{fermions:action4} is
\begin{equation}
\label{fermions:eom1}
	\left[ \hat{\Gamma}^a \tilde{\nabla}_a - \frac{i}{\lad} \Gamma^\flatind{01} \right]
	\theta = 0~,
\end{equation}
where now
\begin{equation}
\label{fermions:tilde.nabla.2}
	\tilde{\nabla}_\alpha = \nabla_\alpha +\frac14 A_{\flatind{ij}\alpha} \tau^\flatind{ij}~.
\end{equation}
Instead of using \eqref{fermions:six.decomp}, it is easier to use the alternative $4+2$ decomposition
\begin{equation}
\label{fermions:six.decomp.2}
	\hat{\Gamma}^\flatind{\alpha} = \Ifour \otimes \gamma^\flatind{\alpha}~,
	\qquad
	\hat{\Gamma}^\flatind{\mu} = \hat{\gamma}^\flatind{\mu} \otimes \gamma^\flatind{01}~.
\end{equation}
Thus, \eqref{fermions:eom1} becomes
\begin{equation}
\label{fermions:eom2}
	\left[ \hat{\gamma}^\alpha \tilde{\nabla}_\alpha
	+ \gamma^\flatind{01} \left( \hat{\gamma}^\mu \nabla_\mu
	-\frac{i}{\lad} \right) \right] \theta = 0~.
\end{equation}
Let $\psi_{\mu s}$ be a doublet of 2-d spinors and $\chi_{ls}$ a 4-d spinor satisfying the following 2-d and 4-d Dirac equations, respectively,
\begin{align}
\label{fermions:psi.sol}
	\hat{\gamma}^\alpha \tilde{\nabla}_\alpha \psi_{\mu s} &=
	s \mu \psi_{\mu s}~, &(s&=\pm1, \mu\geq0)~, \\
\label{fermions:chi.sol}
	\hat{\gamma}^\mu \nabla_\mu  \chi_{ls} &= i s\frac{l+2}{\lad} \chi_{ls}~,
	&(s&=\pm 1, l=0,1,2,\ldots)~.
\end{align}
The $\chi_{ls}$ are just the eigenfunctions of the Dirac operator on the 4-sphere \cite{Camporesi:1995fb}. 
Then, expanding $\theta$ as
\begin{equation}
\label{fermions:theta.dec}
	\theta = \sum_{\mu,l,s,s'} a_{\mu lss'}\, \chi_{ls'} \otimes \psi_{\mu s} ~,
\end{equation}
and using the property $\gamma^\flatind{01} \psi_{\mu s} =\psi_{\mu -s}$, \eqref{fermions:eom2} leads to the following relation for the coefficients,
\begin{equation}
\label{fermions:a.rels}
	s\mu \lad a_{\mu lss'} + i [s'(l+2)-1] a_{\mu l-ss'} =0~.
\end{equation}
For \eqref{fermions:a.rels} to have a non-trivial solution, it is necessary that
\begin{equation}
\label{fermions:mu}
	\mu \lad = \begin{cases} l+1 \qquad& \text{for $s'=1$} \\
		l+3 & \text{for $s'=-1$}~. \end{cases}
\end{equation}
Summarizing, the classical field equation for the fermionic fluctuations have been reduced to the 2-d Dirac equation \eqref{fermions:psi.sol} with the eigenvalues \eqref{fermions:mu}. Notice, however, that $\psi_{\mu s}$ is a doublet of 2-d Dirac spinors, and $\tilde{\nabla}_\alpha$ contains the normal bundle connection term. 

%% file: Fluctuations_WL.tex
\section{Spectrum of operators on circular Wilson loops}
\label{spectrum}

We can verify now the predictions for the spectrum of operators on the circular Wilson loop made in \cite{faraggi:2011bb} by matching the diagonalized fluctuations with the multiplets listed in table~3 of that paper.

\begin{table}[ht]
\begin{center}
\renewcommand{\arraystretch}{1.3}
\begin{tabular}{|c|c|c|}
	\hline
	\multicolumn{3}{|c|}{\textbf{bosons}}\\	\hline
	field              & $m^2\lad^2$ & $(h,n)\times(m,l)$  \\ \hline
	$\eta_l\; (l\geq1)$ & $l(l-1)$    & $(l,0)\times(0,l)$  \\ \hline
	$\zeta_l$          & $(l+3)(l+4)$& $(l+4,0)\times(0,l)$  \\ \hline
	$a_l$              & $(l+2)(l+3)$& $(l+3,0)\times(2,l)$  \\ \hline
        $\chi^\flatind{i}_l$&$(l+1)(l+2)$& $(l+2,1)\times(0,l)$  \\ \hline \hline
	\multicolumn{3}{|c|}{\textbf{fermions}}\\ \hline
	field                   & $m\lad$   & $(h,n)\times(m,l)$ \\ \hline
	$\psi_{l+}$             & $(l+1)$       & $(l+\frac32,\frac12)\times(1,l)$ \\ \hline
	$\psi_{l+}$             & $(l+3)$       & $(l+\frac72,\frac12)\times(1,l)$ \\ \hline
\end{tabular}
\caption{Matching of the bulk fields with multiplets of $OSp(4^\ast|4)$, cf.\ table~3 of \cite{faraggi:2011bb}. The quantum numbers have the following meaning: $h$ is the conformal dimension, $n=0,\frac12,1$ stand for $SO(3)$ singlets, doublets and triplets, respectively, $m=0,1,2$ for scalar, spinor and vector fields on $S^5$, respectively, and $l$ is the $S^5$ angular momentum. In general, $l\geq0$, except for the field $\eta_l$, see the discussion at the end of section~\ref{bosons}.}\label{spectrum:table}
\end{center}
\end{table}

Consider the D5-brane dual to the circular half-BPS Wilson loop in the totally antisymmetric representation. The 10-d space-time geometry is $AdS_5 \times S^5$. To make use of the rotational symmetry of the circular Wilson loop, we write the $AdS_5$ metric as
\begin{equation}
\label{spectrum:AdS5}
	\rmd s^2 = \frac{\lad^2}{z^2} \left( -\rmd t^2 +\rmd z^2 +\rmd x^2 + \rmd\lambda^2 +\lambda^2 \rmd \phi^2\right)~,
\end{equation}
where $z\in(0,\infty)$ is the radial coordinate. The 2-d part of the background D5-brane world volume is given by the embedding
\begin{equation}
\label{spectrum:embedding}
	\phi= i\tau~,\qquad \lambda = \rho~,\qquad z= \sqrt{R^2-\rho^2}~,\qquad (0\leq\rho\leq R)~,
\end{equation}
where $R$ is the radius of the Wilson loop. This embedding describes $AdS_2\subset AdS_5$.
Notice that we have included a factor of $i$ in the equation for $\phi$ in order to formally achieve Lorentzian signature on the world-sheet.

When evaluating the solution \eqref{spectrum:embedding} in the classical action \eqref{review:D5action.bos3}, the result is
\begin{eqnarray}
S_{D5}^{(B)}+I_{bound}^{(B)}&=&-\frac{N}{3\pi^2 \alpha'}2\pi L^2  \sin^3\vt \nonumber \\
&=&-\frac{2N\sqrt{\lambda}}{3\pi} \sin^3\vt.
\end{eqnarray}
In the second line we have written the gravity answer using field theory variables to illustrate that it matches perfectly with the leading result for the expectation value of the $k$-antisymmetric Wilson loop computed using the matrix model \cite{Yamaguchi2006,Hartnoll2006}.

Let us determine the geometric quantities needed for the field equations. First, because the bulk is $AdS_5 \times S^5$, the curvature term in \eqref{bosons:chi.i.eq2} simply contributes a mass term of $-2/\lad^2$. Second, as $AdS_2$ is maximally symmetric, the second fundamental forms $H^\flatind{i}{}_{\alpha\beta}$ must be proportional to the 2-d induced metric, $g_{\alpha\beta}$. But because they are also traceless (the effective string world-sheet is minimal), we conclude that $H^\flatind{i}{}_{\alpha\beta}=0$. Third, an explicit calculation using the formulas in appendix~\ref{embed} shows that the $SO(3)$ gauge fields $A_{\flatind{ij}\alpha}$ vanish identically.\footnote{This last statement depends, obviously, on the choice of the normal vectors. In general, one gets a pure gauge $A_{\flatind{ij}\alpha}$.}
Therefore, the modes of the independent bosonic fields, $\chi^\flatind{j}_l$, $a_l$, $\eta_l$ and $\zeta_l$, satisfy massive Klein-Gordon equations on $AdS_2$,
\begin{equation}
\label{spectrum:KG}
	\left( \nabla^\alpha\nabla_\alpha - m^2\right) \varphi=0~.
\end{equation}
The masses can be read off from \eqref{bosons:chi.i.eq2}, \eqref{bosons:a.mu.eq2}, \eqref{bosons:chi.a.diag1} and \eqref{bosons:chi.a.diag2} and are related to the conformal dimensions of the dual operators by the standard formula
\begin{equation}
\label{spectrum:dim.formula}
	h=\frac12 +\sqrt{\frac14+m^2\lad^2}~.
\end{equation}

For the fermions, the field equation \eqref{fermions:psi.sol} is a massive Dirac equation on $AdS_2$,\footnote{Actually, \eqref{fermions:psi.sol} is a doublet of Dirac equations, but the (symplectic) Majorana condition that still must be imposed makes it equivalent to a single Dirac equation with an unconstrained spinor.}
\begin{equation}
\label{spectrum:Dirac}
	\left(\gamma^\alpha \nabla_\alpha -m\right) \psi= 0~,
\end{equation}
and the (dimensionless) masses $m\lad$ are given by \eqref{fermions:mu}. They are related to the conformal dimenions of the dual operators by
\begin{equation}
\label{spectrum:dim.formula.f}
	h=|m|\lad +\frac12~.
\end{equation}

We present our results in table~\ref{spectrum:table} in a form similar to table~3 of \cite{faraggi:2011bb}. The predictions of the multiplet structure made in that paper are fully confirmed.

%% file: Determinant_Zeta.tex
\section{One-loop effective action}
\label{det}

Having found the full spectrum of excitations of the half-BPS D5-brane in $AdS_5\times S^5$ dual to the circular Wilson loop, we now proceed to compute the corresponding one-loop effective action using $\zeta$ function techniques \cite{Hawking1977, Vassilevich2003}. Eigenfunctions of the Laplace and Dirac operators in maximally symmetric spaces and their associated heat kernels have been extensively studied \cite{Burgess1985, Camporesi1990, Camporesi1991, Camporesi1993, Camporesi1994, Camporesi:1995fb}. We shall follow in spirit the recent calculations of logarithmic corrections to the entropy of black holes in \cite{Banerjee:2010qc, Banerjee2011a}, especially with regard to the treatment of zero modes.

We start by providing a general review of the $\zeta$ function method in section~\ref{zeta}, focussing for simplicity on a single massive scalar field and highlighting the scaling properties of the functional determinant. In section~\ref{det:mode.decomp}, the expansion of the bosonic fields into eigenfunctions on $AdS_2$ and $S^4$ is done explicitly, so that we can proceed with the calculation of the bosonic heat kernels in section~\ref{det:mixed}. The fermionic heat kernels are calculated in section~\ref{det:fermion.kernels}, and the results are put together in section~\ref{det:final}.

\subsection{Computing functional determinants}
\label{zeta}

Let $\Delta S$ denote the 1-loop correction to the effective action for a single, real massive scalar field. It is given by
\begin{equation}
\label{zeta:effective.action}
    \e{-\Delta S} = \int \mathcal{D}\phi\,
    \e{-\frac12 \int \rmd^d x\,\sqrt{\det g}\, \phi\left(-\Box+m^2\right)\phi}~.
\end{equation}
where the functional integration measure is defined by
\begin{equation}
\label{zeta:measure}
    1 = \int \mathcal{D}\phi\,\e{-\frac12 \mu^2 \int \rmd^d x\,\sqrt{\det g}\, \phi^2}~.
\end{equation}
The constant $\mu$ of dimension inverse length is needed for dimensional reasons, because $[\phi]=L^{1-d/2}$, so that $[S]=1$. Formally, the functional integral \eqref{zeta:effective.action} is written as a functional determinant 
\begin{equation}
\label{zeta:det.def}
    \e{-\Delta S} = \left[ \operatorname{Det} \left(-\Box +m^2 \right) \right]^{-1/2}~.
\end{equation}

To give an operational definition to these formal expressions, introduce an orthonormal set of eigenstates of $\Box$ satisfying
\begin{equation}
\label{zeta:eigenfunctions}
    -\Box f_n =\lambda_n f_n~,\qquad
    \int \rmd^d x\,\sqrt{\det g}\, f_n f_m =\delta_{nm}~.
\end{equation}
If the spectrum of $\Box$ is continuous, the sum is to be understood as an integral with the appropriate spectral measure. In this basis, the field $\phi$ can be expanded as
\begin{equation}
\label{zeta:expansion}
    \phi=\sum_n \phi_n f_n~.
\end{equation}
Notice the units $[f_n]=L^{-d/2}$ and $[\phi_n]=L$. The integration measure satisfying \eqref{zeta:measure} is 
\begin{equation}
\label{zeta:D}
    \mathcal{D}\phi = \prod_n\left( \frac{\mu}{\sqrt{2\pi}} \rmd\phi_n\right)~,
\end{equation}
and a short calculation shows that \eqref{zeta:effective.action} gives rise to
\begin{equation}
\label{zeta:trace}
    \Delta S = \frac12 \sum_n \ln \frac{\lambda_n+m^2}{\mu^2}~.
\end{equation}

In our case, the masses are proportional to $1/\lad$, where $\lad$ is the radius of the $AdS_2$ and $S^4$ factors. Hence, we can write 
\begin{equation}
\label{zeta:rescale}
	\lambda_n +m^2 = \frac1{L^2} \left(\tilde{\lambda}_n +\tilde{m}^2 \right)~,
\end{equation}
where the $\tilde{\lambda}_n$ are the eigenvalues of the Laplacian $-\tilde{\Box}$ corresponding to $AdS_2 \times S^4$ with unit radius, and $\tilde{m}$ represent dimensionless numbers. Defining the $\zeta$ function
\begin{equation}
\label{zeta:zeta.def}
    \zeta(s) =\sum_n\left(\tilde{\lambda}_n+\tilde{m}^2\right)^{-s}~,
\end{equation}
\eqref{zeta:trace} can be expressed as
\begin{equation}
\label{zeta:trace.zeta}
    \Delta S = -\frac12 \zeta'(0) - \ln (\mu \lad)\, \zeta(0) = - \ln \left(L/L_0\right)\, \zeta(0)~.
\end{equation}
In the last equation, we have traded the inverse length $\mu$ for a renormalization length scale $L_0$ absorbing also the first term.

In order to study the $\zeta$ function, it is convenient to introduce the heat kernel
\begin{empheq}{align}\label{zeta: definition of kernel}
    K(x,y;t)&=\sum_n \e{-\left(\lambda_n+m^2\right)t}f_n(x)f_n(y)~.
\end{empheq}
Here and henceforth, we have dropped the tilde and implicitly assume unit length $L=1$.
By construction, \eqref{zeta: definition of kernel} satisfies the heat equation
\begin{empheq}{align}
    \left(\frac{\partial}{\partial t}-\Box+m^2\right)K(x,y;t)&=0~,
\end{empheq}
with the initial condition $K(x,y;0)=\delta(x,y)$. Setting $x=y$ and integrating over the manifold gives the trace
\begin{empheq}{alignat=3}
    Y(t)&\equiv\int \rmd^d x\,\sqrt{\det g}\,K(x,x;t)&=\sum_n \e{-\left(\lambda_n+m^2\right)t}~.
\end{empheq}
Then, the $\zeta$ function is related to the integrated heat kernel by the Mellin transform,
\begin{empheq}{align}\label{zeta: Mellin transform}
    \zeta(s)&=\frac{1}{\Gamma(s)}\int_0^{\infty} \rmd t\,t^{s-1}Y(t)~.
\end{empheq}
Notice that since $AdS_2\times S^4$ is non-compact, the $\zeta$ function will diverge; $K(x,x;t)$ is independent of $x$ for a homogeneous space. Thus, $Y(t)$ and $\zeta(s)$ are proportional to the volume of unit $AdS_2\times S^4$, which must be regularized.

We can separate the integral in \eqref{zeta: Mellin transform} into
\begin{empheq}{align}\label{zeta: Mellin convergence}
    \zeta(s)&=\frac{1}{\Gamma(s)}\left(\int_0^{1}\rmd t\,t^{s-1}Y(t)+\int_1^{\infty}\rmd t\,t^{s-1}Y(t)\right)~.
\end{empheq}
The second term converges for any $s$ since $Y(t)\sim \e{-\left(\lambda_0+m^2\right)t}$ for large $t$. On the other hand, it can be shown that $Y(t)$ has the asymptotic expansion
\begin{empheq}{align}
    Y(t)&\cong\sum_{n=0}^{\infty}a_nt^{(n-d)/2}~.
\end{empheq}
as $t\rightarrow0^+$. Substituting this in the first term of \eqref{zeta: Mellin convergence} gives
\begin{empheq}{align}
    \frac{1}{\Gamma(s)}\sum_n\frac{a_n}{s+(n-d)/2}~.
\end{empheq}
This shows that $\zeta(s)$ will have poles at $s=d,d-1,\ldots,1$. The pole at $s=0$, however, is removed by the gamma function. Inverting \eqref{zeta: Mellin transform} gives
\begin{empheq}{align}\label{zeta: inverse Mellin}
    Y(t)&=\frac{1}{2\pi i}\oint \rmd s\,t^{-s}\Gamma(s)\zeta(s)~,
\end{empheq}
where the integration contour encircles all the poles of the integrand. In particular, \eqref{zeta: inverse Mellin} implies that
\begin{empheq}{align}
\label{zeta:result}
    \zeta(0)&=a_d~.
\end{empheq}
Thus, the problem of computing functional determinants is mapped to the problem of computing the $t$ independent coefficient in the asymptotic expansion of the integrated heat kernel.

The above derivation can be extended to higher spin fields with analogous results. Each field has its own heat kernel and thus its own $\zeta$ function. The total effective action is obtained by simply adding the contribution of the integrated heat kernels from all the fields present in the theory. 
In the case of massless fields, special attention must be paid to possible zero modes of the corresponding kinetic operators as they must be excluded from the definition of the heat kernel. This can be done in an elegant fashion by subtracting from the final heat kernel its value for large $t$ \cite{Banerjee:2010qc}. It turns out, however, that the pieces of the heat kernel that would have to be subtracted have cancelled between the contributions from various fields. Moreover, it can be argued that, as far as the logarithmic corrections are concerned, the full heat kernel yields the correct result \cite{Banerjee:2010qc}.  
For the fluctuations of the D5-brane, a further complication stems from the fact that some modes are coupled and must be diagonalized. We shall deal with these issues at due moment.

%% file: Determinant_Modes.tex
\subsection{Mode decomposition for the bosons}
\label{det:mode.decomp}

We want to calculate the one-loop effective action for the bosons in the background of the holographic Wilson loop. Let us start with the action \eqref{bosons:action}. There are two points we have to address before doing the path integral. First, our fields have physical dimensions $[\chi]=[a]=L$. Thus, to obtain the canonical dimensions used in the last subsection, we must absorb a square root of $T_5$ into each field. 

Second, \eqref{bosons:action} involves the metric $\hat{g}_{ab}$ defined in \eqref{bosons:hat.g}, which is $AdS_2\times S^4$, with both factors of radius $\lad$. The fluctuation fields, however, were defined on the background world volume, which has a metric $M_{ab}=g_{ab}+\F_{ab}$, as defined by the Born-Infeld part of the action. This change has an influence on the functional integration measures, which is easily accounted for by a suitable rescaling of the fields. Consider the norms for scalar and vector fields on the background world volume, which are used to define the integration measures,
\begin{equation}
\label{det:norms}
\begin{aligned}
	||\chi||^2 &= \int \rmd^6 \xi \sqrt{\det M_{ab}}\, \chi^2 = \sin^5\vt 
	\int \rmd^6 \xi \sqrt{\det \hat{g}_{ab}}\, \chi^2~, \\
	||a||^2 &= \int \rmd^6 \xi \sqrt{\det M_{ab}}\, M^{ab} a_a a_b = \sin^3\vt 
	\int \rmd^6 \xi \sqrt{\det \hat{g}_{ab}}\, \hat{g}^{ab} a_a a_b~.
\end{aligned}
\end{equation}
The integrals on the right hand sides of \eqref{det:norms} are the norms that are used to define the integral measures for the path integral on a manifold with metric $\hat{g}_{ab}$. Therefore, in order to write the action \eqref{bosons:action} in terms of integration variables with standard measure and canonical units, we must rescale the fields by 
\begin{equation}
\label{det:rescale}
	\chi \to \frac{\chi}{\sqrt{T_5 \sin^5 \vt}}~,\qquad 
	a_a \to \frac{a_a}{\sqrt{T_5 \sin^3 \vt}}~.
\end{equation}
Thus, \eqref{bosons:action} gives rise to the Euclidean action
\begin{align}
\notag
	S^{(B,2)}_{D5,E} &= -\frac{1}{\sin^2 \vt} \int \rmd^6 \xi \sqrt{\det \hat{g}_{ab}} \left[ 
	\frac12 \delta_\flatind{ij} \chi^\flatind{i}  \left( \hat{\nabla}_a \hat{\nabla}^a -\frac2{\lad^2} \right) \chi^\flatind{j}
	+\frac12 \chi^\flatind{5} \left( \hat{\nabla}_a \hat{\nabla}^a +\frac4{\lad^2} \right)  \chi^\flatind{5} \right. \\
\label{det:action}
	&\quad \left. -\frac{2i}{\lad} \chi^\flatind{5} \epsilon^{\alpha\beta} f_{\alpha\beta} 
	-\frac14 f^{\mu\nu} f_{\mu\nu} -\frac12 f^{\mu\alpha} f_{\mu\alpha} -\frac14 f^{\alpha\beta} f_{\alpha\beta} \right]~.
\end{align}
Notice the $i$ in the first term on the second line, which stems from switching the explicit Lorentzian $\epsilon_{\alpha\beta}$ to Euclidean signature. The additional factor $1/\sin^2\vt$ in front can be absorbed by rescaling $\hat{g}_{ab}\to \sin^2\vt\, \hat{g}_{ab}$ giving rise to $AdS_2$ and $S^4$ of radius $\lad \sin\vt$. 

There are two difficulties we have to address in the calculation of the heat kernels. First, there is the gauge invariance, $a_a\to a_a +\partial_a \lambda$. Second, the sector consisting of the gauge field $a_\alpha$ and the scalar $\chi^\flatind{5}$ must be diagonalized.
This problem does not allow us to factorize the heat kernel in a straightforward fashion. Therefore, we choose to do a complete mode expansion of the action into eigenstates on $S^4$ and $AdS_2$, which will also allow us to perform the gauge fixing on a state-by-state basis.

Let us start with the mode expansion of the fields appearing in \eqref{det:action}. Fields that are scalars on $S^4$ can be decomposed into spherical harmonics, such as
\begin{equation}
\label{det:S4.scalar}
	\chi^\flatind{5} = \sum\limits_{l=0}^\infty Y_l(\Omega)\, \chi^\flatind{5}_l(\tau,\sigma)~.
\end{equation}
We do not explicitly write the sum over the minor angular momentum quantum numbers, which are easily accounted for by remembering the degeneracies $D_l(4,0)$ given in \eqref{bosons:ev.0}.

The gauge field $a^\mu$, which is a vector on $S^4$, decomposes into
\begin{equation}
\label{det:S4.vector}
	a^\mu = \sum\limits_{l=1}^\infty  \left[ Y_l^\mu(\Omega)\, a_l(\tau,\sigma)
	+ \sqrt{\frac{\lad^2}{l(l+3)}} \left(\hat{\nabla}^\mu Y_l(\Omega) \right)b_l(\tau,\sigma)\right]~.
\end{equation}
In contrast to the expansion of the gauge-fixed classical field, we have to include the longitudinal modes. The square root factor in the second term is necessary in order for the eigenfunctions multiplying the coefficients $b_l$ to be properly normalized.

For the $AdS_2$ part, we work in the Poincar\'e metric \eqref{appendix: AdS2 metric}. The normalized scalar eigenfunctions of the Laplacian are then given by \eqref{appendix: AdS2 eigenfunctions}, with eigenvalues $-\nabla_\alpha\nabla^\alpha\to \lambda_\nu= (\nu^2+1/4)$. Hence, the $AdS_2$ scalars decompose like
\begin{equation}
\label{det:AdS.scalar}
	\chi^\flatind{5} = \int\limits_{-\infty}^\infty \rmd k \int\limits_0^\infty \rmd \nu\,
	f_{(k,\nu)} (x,y) \, \chi^\flatind{5}_{(k,\nu)}(\Omega)~.
\end{equation}

For the $AdS_2$ vector $a^\alpha$ we have to be more careful \cite{Banerjee:2010qc}. Locally, an eigenfunction of the vector Laplacian can be written as $a_\alpha = \lambda^{-1/2} (\nabla_\alpha f_1 + \epsilon_{\alpha \beta}\nabla^\beta f_2)$, where $f_1$ and $f_2$ are eigenfunctions of the scalar Laplacian with the same eigenvalue $\lambda$. In doing so, we must take care to include a zero mode, which is not a normalizable scalar mode, but which gives rise to a normalizable vector mode. This mode comes from the $\nu=i/2$ case of the scalar eigenfunctions and reads (for unit $\lad$)
\begin{equation}
\label{det:tilde.f}
	\tilde{f}_k (x,y) = \frac1{\sqrt{2\pi|k|}} \e{ikx -|k|y}~.
\end{equation}
The full expansion of the vector $a^\alpha$ reads, therefore,
\begin{align}
\notag
	a^\alpha &= \int\limits_{-\infty}^\infty \rmd k \int\limits_0^\infty \rmd \nu\, \sqrt{\frac{\lad^2}{\nu^2+1/4}} \left[
	\left( \nabla^\alpha f_{(k,\nu)}\right) c_{(k,\nu)} (\Omega)
	+ \left(\epsilon^{\alpha\beta} \nabla_\beta f_{(k,\nu)} \right) d_{(k,\nu)} (\Omega) \right] \\
\label{det:AdS.vector}	
	&\quad	+ \int\limits_{-\infty}^\infty \rmd k\, \left(\nabla^\alpha \tilde{f}_k \right)  \tilde{c}_k(\Omega)~.
\end{align}
One can check that the eigenfunctions in front of the mode coefficients $c_{(k,\nu)}$, $d_{(k,\nu)}$ and $\tilde{c}_k$ are orthonormal with respect to the norm $\int \rmd^2 x\sqrt{\det g_{\alpha\beta}}\, a^\alpha a_\alpha$. Remember that now $\epsilon_{\alpha\beta} \epsilon^{\alpha\beta}=+2$, because we are in Euclidean signature.

After doing the mode expansion in \eqref{det:action}, one obtains
\begin{align}
\notag
	S^{(B,2)}_{D5,E} &=  \frac1{2\lad^2\sin^2\vt} \int\limits_{-\infty}^\infty \rmd k \int\limits_0^\infty \rmd \nu
	\sum\limits_{l=0(1)}^\infty \left\{ \left[ \nu^2 +\frac14 + l(l+3) +2  \right] \delta_\flatind{ij}
	\chi^\flatind{i}_{l(k,\nu)} \chi^\flatind{j}_{l(k,\nu)} \right. \\
\notag
	&\quad
	+\left[ \nu^2 + \frac14 + l(l+3) -4 \right] \left(\chi^\flatind{5}_{l(k,\nu)}\right)^2
	+8i \sqrt{\nu^2+\frac14} \chi^\flatind{5}_{l(k,\nu)} d_{l(k,\nu)}
	+ \left[ \nu^2 + \frac14 + l(l+3) \right] d_{l(k,\nu)}^2 \\
\notag
	&\quad\left.
	+\left[ \nu^2 +\left( l+\frac32 \right)^2 \right] a_{l(k,\nu)}^2
	+\left[ \sqrt{l(l+3)}c_{l(k,\nu)}- \sqrt{\nu^2+\frac14} b_{l(k,\nu)} \right]^2
	\right\} \\
\label{det:action.expanded}
	&\quad + \frac1{2\lad^2\sin^2\vt} \int\limits_{-\infty}^\infty
	\rmd k \sum\limits_{l=0}^\infty l(l+3) \tilde{c}_{l,k}^2~.
\end{align}
The summation over $l$ starts with $0$ for the first two lines, but with $1$ for the third line. The last line is the contribution from the special $AdS_2$ vector modes. 

%% file: Determinant_Kernelbosons_Triplet.tex
\subsection{Bosonic heat kernels}
\label{det:mixed}

\paragraph{Triplet}

The calculation is simplest for the triplet fields $\chi^\flatind{i}$. The contribution of each triplet field to the heat kernel is
\begin{equation}
\label{det:Yi.decomp}
	Y^{\chi^\flatind{i}}(t) = \e{-2\bt}\, Y^s_\unitAds(\bt)\, Y^s_\unitS(\bt)~,
\end{equation}
where $\bt = t/(\lad\sin\vt)^2$, and the heat kernels on $\unitAds$ and $\unitS$ (the hats indicate that these are $AdS_2$ and $S^4$ of unit radii) are given, respectively, by \cite{Banerjee:2010qc}
\begin{equation}
\label{det:Y.AdS}
	Y^s_\unitAds(t) = \frac{V_\unitAds}{2\pi} \e{-t/4} \int\limits_0^\infty \rmd \nu \, \nu
	\tanh(\pi\nu) \e{-\nu^2 t}
\end{equation}
and
\begin{equation}
\label{det:Y.S}
	Y^s_\unitS(t) = \sum\limits_{l=0}^\infty D_l(4,0) \e{-l(l+3)t}
	= \sum\limits_{l=0}^\infty \frac16(l+1)(l+2)(2l+3) \e{-l(l+3)t}~.
\end{equation}
$V_\unitAds=V_{AdS_2}/\lad^2$ denotes the regulated volume of unit $AdS_2$. The superscript $s$ on the heat kernels indicates that they are for scalar fields.

Let us rewrite \eqref{det:Y.S} by completing the square in the exponent, including the value $l=-1$ in the sum (this does not alter the sum) and shifting the summation index by one. This yields
\begin{equation}
\label{det:Y.S.2}
	Y^s_\unitS(t) = -\frac1{12} \e{9 t/4} \left(1+4\frac{\partial}{\partial t} \right)
	\Sigma^s(t)~,
\end{equation}
with
\begin{equation}
\label{det:Ss}
	\Sigma^s(t)= \sum\limits_{l=0}^\infty \left(l+\frac12\right) \e{-(l+1/2)^2 t}~.
\end{equation}

The evaluations of the integral in \eqref{det:Y.AdS} and the infinite sum in \eqref{det:Ss} are carried out in appendix~\ref{calcsigma}. Substituting the results into \eqref{det:Yi.decomp} we obtain
\begin{align}
\notag
	Y^{\chi^\flatind{i}}(t) &= \frac{V_\unitAds}{2\pi}
	\left[-\frac1{12} \left(1+4\partial_\bt\right) \Sigma^s(\bt)\right]
	\left[-\Sigma^s(-\bt)\right] \\
\label{det:Yi.res}
	&= \frac{V_\unitAds}{2\pi} \left( \frac1{12\bt^3} -\frac1{36\bt^2} -\frac1{756} +\cdots \right)~.
\end{align} 

%% file: Determinant_Kernelbosons_Transverse.tex
\paragraph{Transverse gauge modes}

Let us integrate over the transverse modes $a_l(k,\nu)$, where $l\geq1$. From \eqref{det:action.expanded} we can read off the contribution to the heat kernel
\begin{equation}
\label{det:Y.vec}
	Y^{a}(t) = Y^s_\unitAds (\bt) Y^v_\unitS (\bt)~,
\end{equation}
where
\begin{equation}
\label{det:Y.v.S4}
	Y^v_\unitS(t) = \sum\limits_{l=1}^\infty D_{l}(4,1) \e{-(l+1)(l+2)t}
	= \sum\limits_{l=1}^\infty \frac12l(l+3)(2l+3) \e{-(l+1)(l+2)t}~,
\end{equation}
while $Y^s_\unitAds(\bt)$ is the scalar heat kernel \eqref{det:Y.AdS}. The infinite sum in \eqref{det:Y.v.S4} can be re-written as
\begin{equation}
\label{det:Y.v.S4.2}
	Y^v_\unitS(t) = -\e{t/4} \left(\frac{\partial}{\partial t} +\frac94 \right) \Sigma^s(t) +1~,
\end{equation}
where the $1$ can be traced back to a missing $l=0$ summand after shifting the summation index.


The action \eqref{det:action} is invariant under the gauge symmetry $a_\mu\to a_\mu +\frac{1}{\sin\vt}\partial_\mu\lambda$, where the factor $\frac1{\sin\vt}$ in front of the gauge parameter stems from the different rescalings of scalar and vector fields needed to get standard integration measures, as discussed at the beginning of section~\ref{det:mode.decomp}. Expanding also $\lambda$ into modes, this translates into
\begin{equation}
\label{det:inv.modes}
\begin{split}
	b_{l(k,\nu)} &\to b_{l(k,\nu)}+ \frac{\sqrt{l(l+3)}}{\lad\sin\vt} \lambda_{l(k,\nu)}
	\qquad (l\geq1)~, \\
	c_{l(k,\nu)} &\to c_{l(k,\nu)}+ \frac{\sqrt{\nu^2 +1/4}}{\lad\sin\vt} \lambda_{l(k,\nu)}
	\qquad (l\geq0)~.
\end{split}
\end{equation}
Invariance of \eqref{det:action.expanded} under \eqref{det:inv.modes} is immediate upon inspection of the third line in \eqref{det:action.expanded}.

We can now impose a gauge on a mode-by-mode basis. An obvious choice is to fix the coefficients $c_{l(k,\nu)}$, which can be done using Faddeev-Popov. Hence, we must introduce
\begin{equation}
\label{det:FP}
	\delta\left(c_{l(k,\nu)} \right) \frac{\sqrt{\nu^2 +1/4}}{\lad\sin\vt}
\end{equation}
into the functional integral, where the second factor is the Faddeev-Popov determinant. However, performing the integral over $b_{l(k,\nu)}$, we obtain $\frac{\lad\sin\vt}{\sqrt{\nu^2 +1/4}}$, but only for $l\geq1$. Hence, the net result of gauge fixing, the trivial integration over $c_{l(k,\nu)}$ and the integration over $b_{l(k,\nu)}$ is minus the contribution of an $AdS_2$ scalar,
\begin{equation}
\label{det:Y.gf}
	Y^{gf,b,c}(t) = - Y^s_{\widehat{AdS}_2}(\bt)~.
\end{equation}
This compensates the $1$ in \eqref{det:Y.v.S4.2}. Hence, after gauge fixing, the heat kernel for the vector fields $a_\mu$ is
\begin{align}
\notag
	Y^{a_\mu}(t) = Y^a(t) + Y^{gf,b,c}(t)
	&= \frac{V_\unitAds}{2\pi}
	\left[-\left(\frac94+\partial_\bt\right) \Sigma^s(\bt)\right]
	\left[-\Sigma^s(-\bt)\right] \\
\label{det:Ya.res}
	&= \frac{V_\unitAds}{2\pi} \left( \frac1{4\bt^3} -\frac7{12\bt^2} -\frac{19}{1260} +\cdots \right)~.
\end{align} 

%% file: Determinant_Kernelbosons_Mixed.tex
\paragraph{Mixed sector}

To integrate over $\chi^\flatind{5}_{l(k,\nu)}$ and $d_{l(k,\nu)}$, we have to deal with the matrix
\begin{equation}
\label{det:chi.d.mat}
	M= \begin{pmatrix}
	\nu^2 +\frac14 +l(l+3) -4 & 4i \sqrt{\nu^2+\frac14} \\
	4i \sqrt{\nu^2+\frac14} & \nu^2 +\frac14 +l(l+3)
	\end{pmatrix}~.
\end{equation}
Its eigenvalues are
\begin{equation}
\label{det:chi.d.mat.ev}
	(\nu\pm 2i)^2 +\frac14 +l(l+3) +2~,
\end{equation}
but its determinant can also be written in terms of real factors,
\begin{equation}
\label{det:chi.d.mat.det}
	\det M= \left[\nu^2+\frac14 +l(l-1)\right] \left[\nu^2+\frac14 +(l+3)(l+4)\right]~.
\end{equation}
The two factors on the right hand side of \eqref{det:chi.d.mat.det} are precisely what one would expect from the classical spectrum.

It is possible to calculate the heat kernel either from the eigenvalues \eqref{det:chi.d.mat.ev} or the factors in \eqref{det:chi.d.mat.det}. The results of the calculations differ in the scheme dependent divergent terms $1/t^2$ and $1/t$, but we shall perform both calculations, because a similar ambiguity will be encountered for the fermions.

The heat kernel calculation using the eigenvalues \eqref{det:chi.d.mat.ev} is similar to the situation encountered in \cite{Banerjee:2010qc}, and we shall follow the treatment of that paper. The effect of the mixing between the scalar and the gauge field is a complex shift of the $AdS_2$ eigenvalue compared to \eqref{det:Y.AdS}. Hence, the integrated heat kernel for the $\chi^\flatind{5}$ and $d$ integration is
\begin{equation}
\label{det:Y.chi.d.1}
	Y_1^{\chi^\flatind{5},d}(t) = \e{-2\bt} Y^s_\unitS (\bt)
	\left[ 2 Y^s_\unitAds (\bt) +\delta Y^s_\unitAds (\bt)\right]~,
\end{equation}
where
\begin{equation}
\label{det:dY.chi.d}
	\delta Y^s_\unitAds (t) = \frac{V_\unitAds}{2\pi} \e{-t/4}
	\int\limits_0^\infty \rmd \nu\, \nu
	\tanh(\pi \nu) \left[ \e{-(\nu-2i)^2 t} + \e{-(\nu+2i)^2 t} -2 \e{-\nu^2 t} \right]~.
\end{equation}
For the first two terms in the integrand of \eqref{det:dY.chi.d}, we shift the integration variables to $\nu-2i$ and $\nu+2i$, respectively, such as to obtain the same exponent as in the third term. Then, we deform the integral contours such that we have integrals from $-2i$ ($+2i$) to $0$ (staying to the right of the imaginary axis) and from $0$ to $\infty$. The latter cancel against the third term in \eqref{det:dY.chi.d}. Finally, switching the sign of the integration variable in one of the two remaining integrals, they can be combined into
\begin{equation}
\label{det:dY.chi.d.1}
	\delta Y^s_\unitAds(t) = \frac{V_\unitAds}{2\pi} \e{-t/4} \oint \rmd \nu \left(\nu -2i\right)
	\tanh(\pi\nu) \e{-\nu^2 t}~,
\end{equation}
where the integration contour circles clockwise around the poles at $\nu=i/2$ and $\nu=3i/2$.
The residue theorem then yields
\begin{equation}
\label{det:dY.chi.d.2}
	\delta Y^s_\unitAds (t) = - \frac{V_\unitAds}{2\pi} \left( \e{2t} +3 \right)~.
\end{equation}

Putting everything together, we get
\begin{align}
\notag
	Y_1^{\chi^\flatind{5},d}(t) &= Y^s_\unitS(\bt)
	\left[ 2\e{-2\bt} Y^s_\unitAds(\bt) - \frac{V_\unitAds}{2\pi} \left(3 \e{-2\bt} +1 \right) \right] \\
\notag
	&= \frac{V_\unitAds}{2\pi} \left[ -\frac1{12}\left( 1+4\partial_\bt\right) \Sigma^s(\bt) \right]
	\left[-2\Sigma^s(-\bt) -3 \e{\bt/4} - \e{9\bt/4} \right] \\
\label{det:Y.chi.d.1.res}
	&= \frac{V_\unitAds}{2\pi}
	\left( \frac1{6\bt^3} -\frac{13}{18\bt^2} -\frac1{3\bt} -\frac{551}{1890} +\cdots \right)~.
\end{align}
Strictly speaking, we should have removed a zero mode by subtracting the value of the integrated heat kernel at $t=\infty$. One easily finds from \eqref{det:Y.S} that $Y^s_\unitS(\infty) = 1$, so the last term in the brackets on the first line of \eqref{det:Y.chi.d.1.res} contains a zero mode. We shall ignore this for the moment, because the subtraction can be done at the very end \cite{Banerjee:2010qc}.

Let us now consider the alternative choice, namely, we perform the calculation using the factors of the determinant \eqref{det:chi.d.mat.det}. In this case we get
\begin{equation}
\label{det:Y.chi.d.2}
	Y_2^{\chi^\flatind{5},d}(t) = Y^s_\unitAds (\bt) \sum_{l=0}^\infty D_l(4,0)
	\left[ \e{-l(l-1)\bt} +\e{-(l+3)(l+4)\bt} \right]~.
\end{equation}
The infinite sum can be re-written as
\begin{equation}
\label{det:Y.chi.d.sum}
	\sum_{l=0}^\infty D_l(4,0) \left[ \e{-l(l-1) t} +\e{-(l+3)(l+4) t} \right]
	= \frac23 \e{t/4} \left(-\partial_t +\frac{47}4 \right) \Sigma^s(t) +2~,
\end{equation}
where the $2$ stems from extra terms due to shifts of the summation index.
Thus, the final result is
\begin{align}
\notag
	Y_2^{\chi^\flatind{5},d}(t) &= \frac{V_\unitAds}{2\pi}
	\left[ \frac23 \left(\frac{47}4-\partial_\bt \right) \Sigma^s(\bt) +2 \e{-\bt/4} \right]
	\left[-\Sigma^s(-\bt)\right] \\
\label{det:Y.chi.d.2.res}
	&= \frac{V_\unitAds}{2\pi}
	\left( \frac1{6\bt^3} +\frac{35}{18\bt^2} +\frac1{\bt} -\frac{551}{1890} +\cdots \right)~.
\end{align}
As anticipated, the results \eqref{det:Y.chi.d.1.res} and \eqref{det:Y.chi.d.2.res} differ in the scheme dependent $1/t^2$ and $1/t$ terms. It is worth noting that the relevant terms for the final result, that is, the leading $1/t^3$ terms and the constant terms, are identical in both choices. 

%% file: Determinant_Kernelbosons_Special.tex
\paragraph{Special modes}

Finally, let us integrate over the special modes $\tilde{c}$. The $AdS_2$ part of their heat kernel is obtained from the wave functions \eqref{det:tilde.f} as
\begin{equation}
\label{det:K.tc}
	K^{\tilde{c}}_\unitAds (t) = \int\limits_{-\infty}^\infty \rmd k\, (\nabla^\mu \tilde{f}_k)(\nabla_\mu \tilde{f}_k)
	= \int\limits_{-\infty}^\infty \rmd k\, \frac{2 k^2 y^2}{2\pi |k|} \e{-2|k|y} = \frac1{2\pi}~.
\end{equation}
This is independent of $t$, because the special modes are zero modes on $AdS_2$.

Thus, the integrated heat kernel for the special $AdS$ vector modes is
\begin{equation}
\label{det:Y.tc}
	Y^{\tilde{c}}(t) = \frac{V_\unitAds}{2\pi} Y^s_\unitS(\bt) = \frac{V_\unitAds}{2\pi}
	\left( \frac1{6\bt^2} +\frac1{3t} +\frac{29}{90} +\cdots \right)~.
\end{equation}
Note that we have not subtracted the zero mode $l=0$. However, one can recognize that $Y^{\tilde{c}}(t)$ cancels the third term in the brackets on the first line of \eqref{det:Y.chi.d.1.res}, which contains the zero mode from the $(\chi^\flatind{5},d)$ sector, as discussed above. Thus, all bosonic zero modes cancel precisely, and no further subtraction is necessary. 

%% file: Determinant_Kernelbosons_All.tex
\paragraph{All bosonic modes}

Let us put together the results for all bosonic fields, \eqref{det:Yi.res}, \eqref{det:Ya.res}, \eqref{det:Y.chi.d.1.res} [or \eqref{det:Y.chi.d.2.res}] and \eqref{det:Y.tc},
\begin{equation}
\label{det:Y.bosons}
	Y^{bos}(t) = 3 Y^{\chi^\flatind{i}}(t) + Y^{a_\mu}(t) + Y^{\chi^\flatind{5},d}(t)
	+ Y^{\tilde{c}}(t)~.
\end{equation}
Using \eqref{det:Y.chi.d.1.res} for the mixed sector, we obtain
\begin{equation}
\label{det:Y.bos.1}
	Y_1^{bos}(t) = \frac{V_\unitAds}{2\pi}
	\left( \frac2{3\bt^3} -\frac{11}{9\bt^2} +\frac{11}{945} +\cdots \right)~,
\end{equation}
while using \eqref{det:Y.chi.d.2.res} gives rise to
\begin{equation}
\label{det:Y.bos.2}
	Y_2^{bos}(t) = \frac{V_\unitAds}{2\pi}
	\left( \frac2{3\bt^3} +\frac{13}{9\bt^2} +\frac{4}{3\bt} +\frac{11}{945} +\cdots \right)~.
\end{equation}

%% file: Determinant_Kernelfermions.tex
\subsection{Fermionic heat kernels}
\label{det:fermion.kernels}

We have seen in section~\ref{fermions} that there are equivalent ways of writing the 6-d fermionic action that are related to each other by chiral rotations. As is well known \cite{Fujikawa:1979ay}, the fermion integration measure is, in general, not invariant under a chiral rotation in the presence of curvature or gauge fields. To detect whether this is an issue here, let us calculate the fermionic heat kernels corresponding to the actions \eqref{fermions:action4} and \eqref{fermions:action5}, in both cases using a the standard measure for the fermions. We will find that the resulting heat kernels differ in the scheme-dependent $1/t^2$ and $1/t$ terms, but the leading $1/t^3$ term and the constant term are identical, just as we found in the mixed sector of the bosons. This implies that we can safely ignore generic problems with the measure under chiral rotations.
Remember that in \eqref{fermions:action4} and \eqref{fermions:action5} the mass term commutes with either the 2-d or the 4-d part of the kinetic term and anti-commutes with the other. In both cases, the two 6-d spinors in the doublet are not coupled, because $A_{\flatind{ij}\alpha}=0$, and the symplectic Majorana condition \eqref{fermions:majorana.six} reduces the doublet to a single independent Dirac spinor, giving rise to a factor of 2 in the action.

Let us start with \eqref{fermions:action5}. Arguing as for the bosons, we find that $\theta$ must be re-scaled like a scalar, so that \eqref{fermions:action5} gives rise to
\begin{equation}
\label{det:action.fermions}
	S^{(F)}_{D5,E} = \frac1{\sin\vt} \int \rmd^6 \xi \sqrt{\det \hat{g}_{ab}}\, \bar{\theta}
	\left[ \hat{\Gamma}^a \nabla_a
	+\frac{1}{\lad} \hat{\Gamma}^\flatind{6789} \right] \theta~.
\end{equation}
Again, the factor $1/\sin\vt$ in front can be absorbed by rescaling the metric.  
Analogous arguments hold for the action \eqref{fermions:action4}.

Writing  the Dirac operator in the brackets of \eqref{det:action.fermions} as
\begin{equation}
\label{det:dirac1}
	D = \hat{\Gamma}^\mu \nabla_\mu +
	\left( \hat{\Gamma}^\alpha \nabla_\alpha + \frac1{\lad} \hat{\Gamma}^\flatind{6789} \right)~,
\end{equation}
one can verify that the two terms on the right hand side anti-commute. The 4-d Dirac operator on $S^4$, $\hat{\Gamma}^\mu \nabla_\mu$, has eigenvalues $\pm i(l+2)/\lad$ $(l=0,1,2,\ldots)$ with degeneracy $D_l(4,\frac12)= \frac23 (l+1)(l+2)(l+3)$ \cite{Camporesi:1995fb}. The 2-d Dirac operator on AdS$_2$, $\hat{\Gamma}^\alpha \nabla_\alpha
$ has a continuous spectrum $i\lambda/\lad$ $(\lambda\geq0$; the spectral measure can be found in \cite{Camporesi:1995fb, Banerjee:2010qc}). Taking the square of $D$, we get
\begin{equation}
\label{det:dirac1.square}
	D^2 = \left( \hat{\Gamma}^\mu \nabla_\mu \right)^2+
	\left( \hat{\Gamma}^\alpha \nabla_\alpha + \frac1{\lad} \hat{\Gamma}^\flatind{6789} \right)^2~.
\end{equation}
Because $\hat{\Gamma}^\flatind{6789}$ commutes with $\hat{\Gamma}^\alpha \nabla_\alpha$ and has eigenvalues $\pm1$, we obtain the integrated heat kernel as
\begin{equation}
\label{det:Yf1}
	Y^f_1(t) = - Y^f_\unitS(\bt) \left[ 2 Y^f_\unitAds(\bt) + \delta Y^f_\unitAds(\bt) \right]~,
\end{equation}
where
\begin{equation}
\label{det:Yf.S4}
	Y^f_\unitS(t) = -\sum\limits_{l=0}^\infty D_l(4,\textstyle{\frac12}) \e{-(l+2)^2 t}~,
\end{equation}
\begin{equation}
\label{det:Yf.AdS2}
	Y^f_\unitAds(t) = -\frac{V_\unitAds}{2\pi} 2 \int\limits_0^\infty \rmd \lambda\, \lambda
	\coth(\pi\lambda) \e{-\lambda^2 t}
\end{equation}
and
\begin{equation}
\label{det:dYf.AdS2}
	\delta Y^f_\unitAds(t) = -\frac{V_\unitAds}{2\pi} 2 \int\limits_0^\infty \rmd \lambda\, \lambda
	\coth(\pi\lambda)
	\left[ \e{-(\lambda+i)^2 t} + \e{-(\lambda-i)^2 t}  -2 \e{-\lambda^2 t} \right]~.
\end{equation}
The $S^4$ part \eqref{det:Yf.S4} is re-written as
\begin{equation}
\label{det:Yf.S4.2}
	Y^f_\unitS(t) = \frac23 \left(\partial_t +1 \right) \Sigma^f(t)~,
\end{equation}
where we have introduced
\begin{equation}
\label{det:Sf}
	\Sigma^f(t) = \sum\limits_{l=0}^\infty l \e{-l^2 t}~.
\end{equation}
The explicit evaluation of $\Sigma^f$ and the integral in \eqref{det:Yf.AdS2} are relegated  to appendix~\ref{calcsigma}. The expression \eqref{det:dYf.AdS2} is obtained along the lines of the first mixed sector calculation in section~\ref{det:mixed}. One obtaines the contour integral
\begin{equation}
\label{det:dYf.AdS2.2}
	\delta Y^f_\unitAds(t) = \frac{V_\unitAds}{2\pi} 2 \oint\limits_{0^+}^{0^-}
	\rmd \lambda\, (\lambda -i) \coth(\pi\lambda) \e{-\lambda^2 t}~,
\end{equation}
where the contour runs from $0^+$ to $i$ along the right of the imaginary axis and back to $0^-$ along the left. Note that there are no poles inside the contour, and the integrand is regular at $\lambda=i$. However, we cannot close the contour due to the pole at $\lambda=0$, so that the value of the integral must be defined as the principal value (half of the residue value),
\begin{equation}
\label{det:dYf.AdS2.3}
	\delta Y^f_\unitAds(t) = \frac{V_\unitAds}{2\pi} 2 \pi i\, \res_{\lambda=0} \left[(\lambda -i) 	  \coth(\pi\lambda) \e{-\lambda^2 t} \right]  = \frac{V_\unitAds}{2\pi}\, 2~.
\end{equation}
Collecting everything together, we obtain
\begin{align}
\notag
	Y^f_1(t) &= - \frac{V_\unitAds}{2\pi}
	\left[\frac23 \left(\partial_\bt+1\right) \Sigma^f(\bt) \right]
	\left[4 \Sigma^f(-\bt) +2 \right] \\
\label{det:Yf1.res}
	&= - \frac{V_\unitAds}{2\pi}
	\left( \frac{2}{3t^3} -\frac{11}{9t^2} +\frac2{3t} -\frac{271}{3780}+\cdots \right)~.
\end{align}

Starting, instead, with the action \eqref{fermions:action4}, we have the Dirac operator\footnote{In Euclidean signature, the 2-d chirality matrix is $\hat{\Gamma}^\flatind{01} = i\hat{\Gamma}^\flatind{0}\hat{\Gamma}^\flatind{1}$, so that the property $(\hat{\Gamma}^\flatind{01})^2=1$ is maintained.}
\begin{equation}
\label{det:dirac2}
	D = \left( \hat{\Gamma}^\mu \nabla_\mu -\frac{i}{\lad} \hat{\Gamma}^\flatind{01} \right) +
	\hat{\Gamma}^\alpha \nabla_\alpha~.
\end{equation}
Analogous arguments as above lead to the integrated heat kernel
\begin{equation}
\label{det:Yf2}
	Y^f_2(t) = Y^f_\unitAds(\bt) \sum\limits_{l=0}^\infty D_l(4,\textstyle{\frac12})
	\left[ \e{-(l+1)^2 \bt}+ \e{-(l+3)^2 \bt} \right]~.
\end{equation}
After a short calculation, the infinite sum can be re-written as
\begin{equation}
\label{det:Yf.S4.mod}
	\sum\limits_{l=0}^\infty D_l(4,{\textstyle \frac12}) \left[ \e{-(l+1)^2 t}+ \e{-(l+3)^2 t} \right]
	= \frac43 \left(2-\partial_t \right) \Sigma^f(t)~.
\end{equation}
Hence, after substituting the results into \eqref{det:Yf2}, we obtain
\begin{align}
\notag
	Y^f_2(t) &= - \frac{V_\unitAds}{2\pi}
	\left[\frac83 \left(\partial_\bt-2\right) \Sigma^f(\bt) \right]
	\Sigma^f(-\bt)  \\
\label{det:Yf2.res}
	&= - \frac{V_\unitAds}{2\pi}
	\left( \frac{2}{3t^3} +\frac{13}{9t^2} -\frac{271}{3780}+\cdots \right)~.
\end{align}
As already anticipated from the results of the mixed sector bosons, the two ways of calculating the heat kernel lead to results that differ in the scheme-dependent $1/t^2$ and $1/t$ terms, but yields identical results for the leading $1/t^3$ and the constant terms. 

%% file: Determinant_Kernelfinal.tex
\subsection{Combining bosons and fermions}
\label{det:final}

We are now in a position to give the full answer for the heat kernel. As we have two slightly different expressions for the bosons and two for the fermions, there would be four different combinantions. One can readily see that the leading $1/t^3$ term cancels in all of them, and the constant term, which is responsible for the scaling, is always the same. We can, however, make the following nice observation, which indicates that supersymmetry does more than just cancelling the leading term. It appears natural to combine \eqref{det:Y.bos.1} with \eqref{det:Yf1.res}, because the heat kernels of the mixed sector bosons and the fermions were calculated with a shift of the eigenvalues on the AdS$_2$ part. Similarly, we should add \eqref{det:Y.bos.2} and \eqref{det:Yf2.res}, for which the eigenvalue shifts happend on the $S^4$ part. In these combinations, also the $1/t^2$ terms cancel, and we obtain
\begin{align}
\label{det:Y1}
	Y_1(t) &= Y^s_1(t)+ Y^f_1(t) = \frac{V_\unitAds}{2\pi}
	\left( -\frac{2}{3t} +\frac1{12} +\cdots \right)~, \\
\label{det:Y2}
	Y_2(t) &= Y^s_2(t)+ Y^f_2(t) = \frac{V_\unitAds}{2\pi}
	\left( \frac{4}{3t} +\frac1{12} +\cdots \right)~.
\end{align}

It remains to regularize the infinite volume $V_\unitAds$, for which we follow the treatment of \cite{Banerjee2011a} complemented with a field theory prescription due to Polyakov \cite{Polyakov:1980ca}. For the circular Wilson loop, it is appropriate to describe unit AdS$_2$ by the metric
\begin{equation}
\label{det:AdS.metric}
	\rmd s^2 = \rmd \eta^2 + \sinh^2 \eta \, \rmd \phi^2~.
\end{equation} 
To regularize the volume we introduce a cut-off $\eta_0$, so that the regularized volume of AdS$_2$ is $2\pi (\cosh\eta_0 -1)$. In the context of corrections to the entropy of black holes \cite{Banerjee:2010qc} the interpretation of the regularization is as follows.  When substituted in the effective action, the term proportional to $\cosh \eta_0$ gives rise, up to a term that vanishes when $\eta_0\to \infty$, to a divergent contribution $\beta \Delta E$, where $\beta \sim 2\pi \sinh\eta_0$ is the inverse temperature and $\Delta E$ is the shift in the ground state energy due to the introduction of the cut-off. This regularization has a simple interpretation on the field theory side as well. In \cite{Polyakov:1980ca}, Polyakov studied the evaluation of vacuum expectation values of general Wilson loops and determined a divergent term that is proportional to the length of the contour and can be interpreted as the mass renormalization of the test particle traveling around the contour. Either interpretation leads, for the one-loop correction, to
\begin{equation}
\label{det:AdS.volume}
	V_\unitAds=-2\pi~. 
\end{equation} 

Let us now collect the various pieces and give the final result. Using \eqref{zeta:trace.zeta}, \eqref{zeta:result}, \eqref{det:Y1} and \eqref{det:AdS.volume} taking into account also that the appropriate radius of the manifold for canonically normalized fields is $\lad \sin\vt$, as discussed after \eqref{det:action}, we find for the one-loop effective action
\begin{equation}
\label{det:final.result}
	\Delta S = \frac1{12} \ln \frac{\lad \sin\vt}{L_0}~.
\end{equation}

%% file: Conclusions.tex
\section{Conclusions}
\label{sec:conclusions}

In this paper, we have explicitly treated the D5-brane configuration dual to the half-BPS circular Wilson loop in the totally antisymmetric representation. We have derived the fluctuations in both, the bosonic and the fermionic sectors. We have also verified that the excitations fall precisely in the expected supermultiplets of $OSp(4^*|4)$. Lastly, we computed the one-loop determinants and provided an answer for the effective action at the one-loop level.

Our work is largely motivated by the applications to the Wilson loops and the potential to take the correspondence beyond the classical ground state by incorporating quantum corrections. This provides a step towards being able to directly compare one-loop corrections from the field theory (Matrix model) and gravity (D-brane) sides. More generally, our work represents a systematic exploration of the various issues that can arise during the quantization of extended objects in the context of the AdS/CFT correspondence. We have encountered and resolved various ambiguities and in the process shed some light on the type of issues that need to be resolved if a coherent quantization of extended objects in curved backgrounds is to be achieved. For example, we hope to have fully clarified the, at times \emph{ad hoc}, process of computing the action for the quadratic fluctuations by explicitly highlighting the differential geometric nature of the fluctuations. We also resolved various technical issues in the computation of the heat kernel for fermions and showed a natural way to determine a scheme. More importantly, at least in our example, we witness that the role of supersymmetry seems to go beyond the expected cancellation of the leading divergence.

There are a few very interesting problems that follow naturally from our work, and we finish by highlighting some of them.

\begin{itemize}

\item A natural direction is the calculation and comparison with the matrix model. We hope to report on this interesting issue in an upcoming publication. The task at hand, although conceptually clear, is plagued with many technical issues. Some of these issues are generic to the whole program of comparing expectation values of operators in the field theory and in the gravity dual. We mentioned in the introduction that, even in the apparently simple case of the Wilson loop in the fundamental representation, an agreement has not been found  \cite{Forste:1999qn, Drukker:2000ep, Sakaguchi:2007ea, Kruczenski:2008zk}. Hopefully, the extra knob that constitutes the representation might lead to some simplifications.

\item In this paper we did not discuss the field theory dual beyond the mere mentioning of the role as half BPS Wilson loops. An important interpretation is provided by the D5-branes as a dual to a one-dimensional defect CFT and has been quoted in recent works as a model for interesting condensed matter phenomena related to quantum impurity models \cite{Sachdev:2010uj, Mueck:2010ja, Harrison:2011fs}. A similar interpretation of D6-branes as dual descriptions of fermionic impurities in $\mathcal{N}=6$ supersymmetric Chern-Simons-matter theories in $2+1$ dimensions has been advanced in \cite{Benincasa:2011}. In such contexts, uncovering the precise role of the spectrum of excitations should lead to a deeper understanding of the interactions of the system.

\item More generally, our paper provides a first solid step in the direction of analyzing extended objects at the quantum level. It seems that the analysis of \emph{conformal} branes, that is branes whose world volume contains $AdS$ factors, avoids dealing with the daunting issues encountered in the quantization of extended objects in asymptotically flat spacetimes. We plan to pursue this analysis in the future.

\item Recently, Sen and collaborators have studied corrections to the entropy of various black hole configurations using techniques similar to those utilized here. The key technical fact that the near horizon geometry of various black holes contains $AdS$ factors seems to provide a tantalizing playground for our methods. We hope that understanding the quantization of such structures at a deeper level might help clarify difficult issues in black hole physics.

\end{itemize}

\section*{Acknowledgments}
We are grateful to A.~Tirziu for collaboration in relevant matters and comments. We are also thankful M.~Kruczenski, L.~Martucci and A.~Ramallo for various comments. A.F. is thankful to Fulbright-CONICYT. The work of A.F.~and L.A.P.Z.~ is partially supported by Department of Energy under grant DE-FG02-95ER40899 to the University of Michigan. W.M.~acknowledges partial support by the INFN research initiative TV12.

%% file: Appendix_Conventions.tex
\section{Conventions}
\label{convs}

We summarize the conventions used throughout the paper. The 10-dimensional curved coordinates are denoted by Latin indices from the middle of the alphabet, $m,n=0,\ldots,9$. Latin indices from the beginning of the alphabet, $a,b=0,\ldots,5$, denote generic coordinates of the D5-branes. Greek indices from the beginning of the alphabet, $\alpha,\beta=\tau,\rho$ are used for the coordinates of the effective string embedded in the a$AdS_5$ part of the background geometry. Greek indices from the middle of the alphabet, $\mu,\nu=6,\ldots,9$, denote the coordinates of the $S^4$ part of the D5-brane world volume. The corresponding flat indices are underlined. In contrast to \cite{martucci:2005rb}, the Levi-Civita symbols $\epsilon_{a_1\ldots a_n}$ are \emph{tensors}, \ie they include the appropriate factors of $\sqrt{|\det g|}$. With the exception of section~\ref{det}, we assume Lorentzian signature for the 2-d part of the world sheet, which implies $\epsilon_{\alpha\beta}\epsilon^{\alpha\beta}=-2$.

%% file: Appendix_Embeddings.tex
\section{Geometry of emdedded manifolds}
\label{embed}

To describe the embedding of the D5-brane world volume in the bulk, we shall use the structure equations of embedded manifolds 
\cite{Eisenhart}. Deviating from the general notation of the main text and of appendix~\ref{convs}, we shall denote with Latin indices $m,n,\ldots$ the curved bulk coordinates and with Greek indices $\alpha,\beta,\ldots$ the world-volume coordinates. Latin indices $i,j$ are used for the directions normal to the world-volume. The corresponding flat indices are underlined. 

A $d$-dimensional Riemannian manifold $\mathbb{M}$ embedded in a $\tilde{d}$-dimensional Riemannian manifold 
$\tilde{\mathbb{M}}$ ($d<\tilde{d}$) is described by $\tilde{d}$ differentiable functions $x^m$
($m=1\ldots \tilde{d}$) of $d$ variables $\xi^\alpha$ ($\alpha =1\ldots d$). 
The $\xi^\alpha$ are coordinates on $\mathbb{M}$ (the world volume), whereas $x^m(\xi)$ specify the location in
$\tilde{\mathbb{M}}$ (the bulk). 
The tangent vectors to the world volume are given by $x^m_\alpha(\xi) \equiv \partial_\alpha x^m(\xi)$. They provide the pull-back of any bulk quantity onto the world volume. For example, the induced metric is 
\begin{equation}
\label{embed:g}
	g_{\alpha\beta}= x^m_\alpha x^n_\beta\, g_{mn}~. 
\end{equation}
In addition, there are $d_\perp=\tilde{d}-d$ normal vectors $N^m_\flatind{i}$, $\flatind{i}=1,\ldots,d_\perp$. Together with the $x^m_\alpha$, they satisfy the orthogonality and completeness relations 
\begin{equation}
\label{embed:ortho}
	N_\flatind{i}^m x_\alpha^n\, g_{mn} =0~,\qquad
	N_\flatind{i}^m N_\flatind{j}^n g_{mn} = \delta_\flatind{ij}~,\qquad
	g^{\alpha\beta} x_\alpha^m x_\beta^n + \delta^\flatind{ij} N_\flatind{i}^m N_\flatind{j}^n = g^{mn}~. 
\end{equation}
We shall adopt a covariant notation raising and lowering indices with the appropriate metric tensors. The freedom of choice of the normal vectors gives rise to a group $O(n)$ of local rotations of the normal frame.  

The geometric structure of the embedding is determined, in addition to the intrisic geometric quantities, by the second fundamental form $H^\flatind{i}{}_{\alpha\beta}$, which describes the extrinsic curvature, and the gauge connection in the normal bundle, 
$A^\flatind{ij}{}_\alpha = -A^\flatind{ji}{}_\alpha$. They are determined by the equations of Gauss and Weingarten, respectively,
\begin{align}
\label{embed:gauss1}
	\nabla_\alpha x^m_\beta \equiv \partial_\alpha x^m_\beta + \Gamma^m{}_{np} x^n_\alpha x^p_\beta -
	\Gamma^\gamma{}_{\alpha\beta} x^m_\gamma &= H^\flatind{i}{}_{\alpha\beta} N_\flatind{i}^m~,\\
\label{embed:weingarten}
	\nabla_\alpha N^m_\flatind{i} \equiv \partial_\alpha N^m_\flatind{i} + \Gamma^m{}_{np} x^n_\alpha N^p_\flatind{i} 
	- A^\flatind{j}{}_{\flatind{i}\alpha} N^m_\flatind{j} & = - H_{\flatind{i}\alpha}{}^\beta x^m_\beta~.
\end{align}
As is evident here, by using the appropriate connections, $\nabla_\alpha$ denotes the covariant derivative with respect to all indices. 
The integrability conditions of the differential equations \eqref{embed:gauss1} and \eqref{embed:weingarten} are the equations of
Gauss, Codazzi and Ricci, which are, respectively,
\begin{align}
\label{embed:gauss2} 
	R_{mnpq} x^m_\alpha x^n_\beta x^p_\gamma x^q_\delta &= R_{\alpha\beta\gamma\delta} 
	+ H^\flatind{i}{}_{\alpha\delta} H_{\flatind{i}\beta\gamma} - H^\flatind{i}{}_{\alpha\gamma} H_{\flatind{i}\beta\delta}~,\\
\label{embed:codazzi}
	R_{mnpq} x^m_\alpha x^n_\beta N^p_\flatind{i} x^q_\gamma &= 
	\nabla_\alpha H_{\flatind{i}\beta\gamma} - \nabla_\beta H_{\flatind{i}\alpha\gamma}~,\\
\label{embed:ricci}
	R_{mnpq} x^m_\alpha x^n_\beta N^p_\flatind{i} N^q_\flatind{j} &= F_{\flatind{ij}\alpha\beta}
	- H_{\flatind{i}\alpha}{}^\gamma H_{\flatind{j}\gamma\beta} + H_{\flatind{i}\beta}{}^\gamma H_{\flatind{j}\gamma\alpha}~,
\end{align}
where $F_{\flatind{ij}\alpha\beta}$ is the field strength in the normal bundle,
\begin{equation}
\label{embed:F}
	F_{\flatind{ij}\alpha\beta} = \partial_\alpha A_{\flatind{ij}\beta} - \partial_\beta A_{\flatind{ij}\alpha} 
	+ A_{\flatind{ik}\alpha} A^\flatind{k}{}_{\flatind{j}\beta} - A_{\flatind{ik}\beta} A^\flatind{k}{}_{\flatind{j}\alpha}~.
\end{equation}
As mentioned before, the covariant derivative in \eqref{embed:codazzi} contains also the connections $A^\flatind{j}{}_{\flatind{i}\alpha}$.

Let us derive the expression for the pull-back of the spinor bulk covariant derivative on the world volume of the brane, which is needed in section~\ref{fermions},
\begin{equation}
\label{embed:D.alpha}
	x_\alpha^m \nabla_m = x_\alpha^m \left( \partial_m +\frac14 \omega_m{}^\flatind{np} \Gamma_\flatind{np} \right)~.
\end{equation}
The bulk spin connections can be obtained by 
\begin{equation}
\label{embed:omega}
	\omega_m{}^\flatind{np} = -e_q^\flatind{p} \left( \partial_m e^{q\flatind{n}} +\Gamma^q{}_{mp} e^{p\flatind{n}} \right)~,
\end{equation}
and similarly for the world volume spin connections. 

Let us pick a local frame adapted to the embedding, 
\begin{equation}
\label{embed:local.frame}
	e^m_\flatind{n} = \begin{cases} 
		x^m_\alpha e^\alpha_\flatind{\alpha} &\text{for $\flatind{n}=\flatind{\alpha}$,} \\
		N^m_\flatind{i} &\text{for $\flatind{n}=\flatind{i}$.} 
	\end{cases}
\end{equation}
Then, using \eqref{embed:gauss1} and \eqref {embed:weingarten}, it is straightforward to show that
\begin{equation}
\label{embed:pb1}
	x_\alpha^m \left( \partial_m e^{q\flatind{n}} +\Gamma^q{}_{mp} e^{p\flatind{n}} \right) = \begin{cases} 
		H^\flatind{i}_{\alpha\beta} N^q_\flatind{i} e^\beta_\flatind{\alpha} + \omega_{\alpha\flatind{\beta\alpha}}
		e^{\beta\flatind{\beta}} x_\beta^q & \text{for $\flatind{n}=\flatind{\alpha}$,}\\
		-H_{\flatind{i}\alpha}{}^\beta x^q_\beta + A^\flatind{j}{}_{\flatind{i}\alpha} N^q_\flatind{j} 
		&\text{for $\flatind{n}=\flatind{i}$.} 
	\end{cases}
\end{equation}
Hence, one finds for the pull-back of the bulk spin connections 
\begin{equation}
\label{embed:pb.spin.conn}
	x_\alpha^m \omega_{m\flatind{\alpha\beta}} = \omega_{\alpha\flatind{\alpha\beta}}~,\qquad 
	x_\alpha^m \omega_{m\flatind{\alpha i}} = -H_{\flatind{i}\alpha\beta} e^\beta_\flatind{\alpha}~,\qquad
	x_\alpha^m \omega_{m\flatind{ij}} = A_{\flatind{ij}\alpha}~.
\end{equation}
Consequently, \eqref{embed:D.alpha} becomes
\begin{equation}
\label{embed:D}
	x_\alpha^m \nabla_m = \nabla_\alpha -\frac12 H_{\flatind{i}\alpha\beta} \Gamma^\beta\Gamma^\flatind{i} 
	+\frac14 A_{\flatind{ij}\alpha} \Gamma^\flatind{ij}~.
\end{equation}

%% file: Appendix_AdS2.tex
\section{Scalar heat kernel on $AdS_2$}\label{appendix: AdS2 Poincare}
Here we show an explicit derivation of the scalar heat kernel on $AdS_2$ using Poincar\'e coordinates and verify that it coincides with the calculation done in global coordinates \cite{Banerjee:2010qc}.

We begin by finding the eigenfunctions and eigenvalues. Consider the $AdS_2$ metric in Poincar\'e coordinates
\begin{empheq}{align}\label{appendix: AdS2 metric}
    ds^2&=\frac{dx^2+dy^2}{y^2}~.
\end{empheq}
The Laplacian reads
\begin{empheq}{align}
    \Box&=y^2\left(\partial^2_x+\partial^2_y\right)~.
\end{empheq}
Assuming a dependence of the form $e^{ikx}$, the spectral problem becomes
\begin{empheq}{align}
    -y^2\left(\partial^2_y-k^2\right)\phi_{(k,\nu)}(y)&=\left(\nu^2+\frac{1}{4}\right)\phi_{(k,\nu)}(y)~,
\end{empheq}
where we have written the eigenvalues as $\nu^2+1/4$.

The two independent solutions to this equation are
\begin{empheq}{alignat=3}
    \phi^{(1)}_{(k,\nu)}(y)&=\sqrt{y}L_{i\nu}\left(|k|y\right)
    &\qquad&\textrm{and}&\qquad
    \phi^{(2)}_{(k,\nu)}(y)&=\sqrt{y}K_{i\nu}\left(|k|y\right)
\end{empheq}
where
\begin{empheq}{align}
    L_{\mu}(z)&=\frac{i\pi}{2}\frac{I_{-\mu}(z)+I_{\mu}(z)}{\sin(\mu\pi)}
    \\
    K_{\mu}(z)&=\frac{\pi}{2}\frac{I_{-\mu}(z)-I_{\mu}(z)}{\sin(\mu\pi)}
\end{empheq}
and $I_{\alpha}$ is the modified Bessel function of the first kind. Of course, $K_{\alpha}$ is the usual modified Bessel function of the second kind. It is better to consider $L_{\alpha}$ and $K_{\alpha}$ (as opposed to $I_{\alpha}$ and $K_{\alpha}$) as independent solutions since they are both real when the order is imaginary and the argument real.

If $\nu$ is purely imaginary, both solutions fail to be square integrable. For real $\nu$, the asymptotic behavior as $y\rightarrow0^+$ is
\begin{empheq}{align}
    L_{i\nu}(y)&=\sqrt{\frac{\pi}{\nu\sinh(\nu\pi)}}\left[\nu\cos\left(\nu\ln(y/2)-c_{\nu}\right)+O(y^2)\right]~,
    \\
    K_{i\nu}(y)&=-\sqrt{\frac{\pi}{\nu\sinh(\nu\pi)}}\left[\nu\sin\left(\nu\ln(y/2)-c_{\nu}\right)+O(y^2)\right]~,
\end{empheq}
where $c_{\nu}$ is a constant, and
\begin{empheq}{align}
    L_{i\nu}(y)&=\frac{1}{\sinh(\nu\pi)}\sqrt{\frac{\pi}{2y}}e^{y}\left[1+O\left(\frac{1}{y}\right)\right]~,
    \\
    K_{i\nu}(y)&=\sqrt{\frac{\pi}{2y}}e^{-y}\left[1+O\left(\frac{1}{y}\right)\right]~,
\end{empheq}
when $y\rightarrow\infty$. From this we see that both solutions vanish as we approach the boundary $y=0$, but only $\phi^{(2)}_{(k,\nu)}$ vanishes as $y\rightarrow\infty$. In other words, only $\phi^{(2)}_{(k,\nu)}$ is square integrable.

The relation (see Kontorovich-Lebedev transform)
\begin{empheq}{align}
    \int_0^{\infty}dy\,\frac{K_{i\mu}(y)K_{i\nu}(y)}{y}&=\frac{\pi^2}{2\mu\sinh(\pi\mu)}\delta(\mu-\nu)~,
\end{empheq}
sets the normalization of the eigenfunctions as
\begin{empheq}{align}\label{appendix: AdS2 eigenfunctions}
    f_{(k,\nu)}(x,y)&=\frac{1}{\sqrt{\pi^3}}\sqrt{\nu\sinh\left(\pi\nu\right)}e^{ikx}\sqrt{y}K_{i\nu}\left(|k|y\right)~,
\end{empheq}
where $k\in\mathbb{R}$ and $\nu\geq0$.

Now, the diagonal heat kernel is
\begin{empheq}{align}
    K\left((x,y),(x,y);t\right))&=\int dkd\nu\,e^{\left(\nu^2+\frac{1}{4}\right)t}f_{(k,\nu)}^*(x,y)f_{(k,\nu)}(x,y)
\end{empheq}
Using the above eigenfunctions this is
\begin{empheq}{align}
    K\left((x,y),(x,y);t\right))&=\frac{1}{\pi^3}\int_0^{\infty} d\nu\,e^{-\left(\nu^2+\frac{1}{4}\right)t}\nu\sinh(\nu\pi)\int_{-\infty}^{\infty}dk\,yK_{i\nu}(|k|y)^2
    \\
    &=\frac{2}{\pi^3}\int_0^{\infty} d\nu\,e^{-\left(\nu^2+\frac{1}{4}\right)t}\nu\sinh(\nu\pi)\int_0^{\infty}dk\,K_{i\nu}(k)^2
\end{empheq}
This does not depend on $y$, as expected. The norm of the modified Bessel function is
\begin{empheq}{align}
    \int_0^{\infty}dx\,K_{i\nu}(x)^2&=\frac{\pi}{4}\Gamma\left(\frac{1}{2}+i\nu\right)\Gamma\left(\frac{1}{2}-i\nu\right)
    \\
    &=\frac{\pi^2}{4\cosh(\pi\nu)}
\end{empheq}
Therefore,
\begin{empheq}{align}
    K\left((x,y),(x,y);t\right))&=\frac{1}{2\pi R^2}\int_0^{\infty} d\nu\,e^{-\left(\nu^2+\frac{1}{4}\right)t}\nu\tanh(\nu\pi)
\end{empheq}
This is the same expression one gets when working with global coordinates on the disk. 

%% file: Appendix_Calcsigma.tex
\section{Integrals and infinite sums}
\label{calcsigma}

We will perform here the evaluation of the integrals and infinite sums needed for the heat kernel calculations of bosons and fermions in section~\ref{det}.

For the bosons, let us start with the infinite sum \eqref{det:Ss},
\begin{equation}
\label{calcsigma:Ss}
	\Sigma^s(t) = \sum\limits_{l=0}^\infty \left(l+\frac12\right) \e{-(l+1/2)^2 t}~.
\end{equation}
Converting the sum into a contour integral that picks up suitable poles, as outlined in [2], one obtains
\begin{equation}
\label{calcsigma:Ss.int}
	\Sigma^s(t) = \im \int\limits_0^{\e{i\kappa}\infty} \rmd \nu \, \nu \tan(\pi\nu) \e{-\nu^2 t}~.
\end{equation}
Here, $0<\kappa\ll 1$, so that $\im \nu>0$ in the integrand. Now, we write $\tan(\pi\nu) = i \tanh(-i\pi \nu)$ and expand the $\tanh$ as 
\begin{equation}
\label{calcsigma:tanh.exp} 
	\tanh(\pi\nu) = 1 - 2\sum\limits_{k=1}^\infty (-1)^{k+1} \e{-2\pi\nu k}
\end{equation}
to obtain
\begin{equation}
\label{calcsigma:tan.exp} 
	\tan(\pi\nu) = i\left[ 1 - 2\sum\limits_{k=1}^\infty (-1)^{k+1} \e{2\pi i\nu k}\right]~.
\end{equation}
The integral in \eqref{calcsigma:Ss.int} can be done exactly for the first term of the expansion, while in the remaining terms we expand $\e{-\nu^2 t}$ as a power series in $t$, integrate and perform the summation over $k$. The result is [cf.\ (2.18) of \cite{Banerjee:2010qc}] 
\begin{align}
\notag
	\Sigma^s(t) &= \frac1{2t} +2 \sum\limits_{n=0}^\infty \frac{(2n+1)!}{n!(2\pi)^{2n+2}} t^n 
	\left(1-2^{-2n-1}\right) \zeta(2n+2) \\
\notag 
	&= \frac1{2t} +\frac12 \sum\limits_{n=1}^\infty \frac{t^{n-1}}{n!} \left(1-2^{1-2n}\right) 
	\left|B_{2n}\right| \\
\label{calcsigma:Ss.res}
	&= \frac1{2t} + \frac1{24} +\frac7{960}t + \frac{31}{16128} t^2 + \Order(t^3)~.
\end{align}
On the first line, $\zeta(s)$ denotes the Riemann zeta function, which we expressed in terms of the Bernoulli numbers $B_{2n}$ on the second line.

Consider now the integral in \eqref{det:Y.AdS}. In analogy with the calculation above, we expand the $\tanh$ using \eqref{calcsigma:tanh.exp} such that the leading term is captured by the integral over the first term of the expansion. For the remaining terms, expand $\e{-\nu^2 t}$ as a power series in $t$, integrate and perform the summation over $k$. The result is [cf.\ (2.15) of \cite{Banerjee:2010qc}]
\begin{equation}
\label{calcsigma:Is.res}
	\int\limits_0^\infty \rmd \nu\, \nu \tanh(\pi\nu) \e{-\nu^2 t} 
	=  \frac1{2t} -\frac12 \sum\limits_{n=1}^\infty \frac{(-t)^{n-1}}{n!} \left(1-2^{1-2n}\right) 
	\left|B_{2n}\right| 
	= - \Sigma^s(-t)~.
\end{equation}
Again, we have expressed the Riemann zeta functions in terms of Bernouuli numbers, and the last equality results from a direct comparison with the second line of \eqref{calcsigma:Ss.res}.

Similar calculations must be done for the fermion contributions. Consider the infinite sum
\eqref{det:Sf}
\begin{equation}
\label{calcsigma:Sf}
	\Sigma^f(t) = \sum\limits_{l=0}^\infty l \e{-l^2 t}~.
\end{equation}
Converting the sum into a contour integral, one obtains
\begin{equation}
\label{calcsigma:Sf.int}
	\Sigma^f(t) = - \im \int\limits_0^{\e{i\kappa}\infty} \rmd \nu \, \nu \cot(\pi\nu) \e{-\nu^2 t}~.
\end{equation}
Write $\cot(\pi\nu) = -i \coth(-i\pi \nu)$ and expand the $\coth$ as 
\begin{equation}
\label{calcsigma:coth.exp} 
	\coth(\pi\nu) = 1 + 2\sum\limits_{k=1}^\infty \e{-2\pi \nu k}
\end{equation}
to obtain
\begin{equation}
\label{calcsigma:cot.exp} 
	\cot(\pi\nu) = -i\left[ 1 + 2\sum\limits_{k=1}^\infty \e{2\pi i\nu k}\right]~.
\end{equation}

Continuing as for $\Sigma^s(t)$, we obtain [cf.\ (3.3.16) of \cite{Banerjee:2010qc}]
\begin{align}
\notag
	\Sigma^f(t) &= 
	\frac1{2t} -\frac12 \sum\limits_{n=1}^\infty \frac{t^{n-1}}{n!} \left|B_{2n}\right| \\
\label{calcsigma:Sf.res}
	&= \frac1{2t} - \frac1{12} -\frac1{120}t - \frac{1}{504} t^2 + \Order(t^3)~.
\end{align}

Finally, an analogous calculation for the integral in \eqref{det:Yf.AdS2} yields
\begin{equation}
\label{calcsigma:If.res}
	\int\limits_0^\infty \rmd \nu\, \nu \coth(\pi\nu) \e{-\nu^2 t} 
	=  \frac1{2t} +\frac12 \sum\limits_{n=1}^\infty \frac{(-t)^{n-1}}{n!} \left|B_{2n}\right|
	= - \Sigma^f(-t)~.
\end{equation}